\documentclass[12pt,a4paper]{article}
\usepackage{a4,epsf,here}
\usepackage{amsmath,amsfonts,amscd,amssymb}
\def\MA{{\rm M}_{\rm A}}
\def\trMA{{\rm tr}\,\MA}
\def\trM{{\rm tr\,M}}

\def\be{\begin{equation}}
\def\ee{\end{equation}}
\def\bea{\begin{eqnarray}}
\def\eea{\end{eqnarray}}
\def\eq#1{(\ref{#1})}
\def\bs{\bigskip}
\def\ms{\medskip}
\def\papa#1#2{\frac{\partial#1}{\partial#2}}
\def\fig#1{Fig.~\ref{#1}}
\def\Sec#1{Sec.~\ref{#1}}
\def\App#1{App.~\ref{#1}}
\def\etal{{\it et al}}
\def\siml{\,\hbox{\kern.1em \lower.6ex \hbox{$\sim$} \kern-1.12em
          \raise.6ex \hbox{$<$} }}
\newcommand{\Figurebb}[9]{
\begin{figure}[H]
\begin{center}
\leavevmode
\epsfysize=#7cm
\epsfbox[#2 #3 #4 #5]{#6}
\par
\parbox{#8cm}{
\caption[figure]{\renewcommand{\baselinestretch}{0.8} \small
                                           \hspace{-0.3truecm}#9}
\label{#1}}
\end{center}
\end{figure}
}

\begin{document}

\baselineskip 14pt

\centerline{\bf \Large Transcritical bifurcations}

\bs

\centerline{\bf \Large in non-integrable Hamiltonian systems}

\bs
\bs

\centerline{\bf Matthias Brack$^1$ and Kaori Tanaka$^{1,2}$}

\bs

\centerline{\it $^1$Institute for Theoretical Physics, University of
Regensburg, D-93040 Regensburg, Germany}

\centerline{\it $^2$Department of Physics and Engineering Physics,
University of Saskatchewan,}

\centerline{\it Saskatoon, SK, Canada S7N 5E2}

\ms

\centerline{\today}

\bs
\ms

\centerline{\bf \large Abstract}

\ms

\baselineskip 10pt
\small{
\noindent
We report on transcritical bifurcations of periodic orbits in 
non-integrable two-dimensional Hamiltonian systems. We discuss
their existence criteria and some of their properties using a 
recent mathematical description of transcritical bifurcations 
in families of symplectic maps. We then present numerical
examples of transcritical bifurcations in a class of generalized
H\'enon-Heiles Hamiltonians and illustrate their stabilities and
unfoldings under various perturbations of the Hamiltonians. 
We demonstrate that for Hamiltonians containing straight-line 
librating orbits, the transcritical bifurcation of these orbits is 
the typical case which occurs also in the absence of any discrete
symmetries, while their isochronous pitchfork bifurcation is an 
exception. We determine the normal forms of both types of bifurcations 
and derive the uniform approximation required to include 
transcritically bifurcating orbits in the semiclassical trace 
formula for the density of states of the quantum Hamiltonian.
We compute the coarse-grained density of states in a specific
example both semiclassically and quantum mechanically and
find excellent agreement of the results.
}

\baselineskip 14pt


\section{Introduction}

The transcritical bifurcation (TCB), in which a pair of stable and 
unstable fixed points of a map exchange their stabilities, is a 
well-known phenomenon in one-dimensional non-Hamiltonian systems. 
A simple example occurs in the quadratic logistic map (see, e.g., 
\cite{lili}): 
\be
x_{n+1} = r\,x_n(1-x_n)\,,
\label{quamap}
\ee
where $\{x_n\}$ are arbitrary real numbers and $r$ is the control
parameter. This map has -- amongst others -- two fixed points 
$x_1^*=0$ and $x_2^*=1-1/r$ which exchange their stabilities at 
the critical value $r=1$. For $r<1$, $x_1^*$ is stable and $x_2^*$ 
is unstable, whereas the inverse is true for $r>1$. [Note that in 
many textbooks discussing the quadratic map, this bifurcation is 
not mentioned as the values of the variable $x$ are usually 
confined to be non-negative, while $x_2^*<0$ for $r<1$.] The TCB 
occurs in many maps used to describe growth or population phenomena 
(see \cite{hogr} for a recent example). TCBs have also been 
reported to occur in various time-dependent model systems 
\cite{lepe,drvi,moza,hwan,koku,bad1,bad2} and shown, e.g., to be 
involved in synchronization mechanisms \cite{moza,hwan}. In 
\cite{bad1,bad2}, TCBs have been found to play a crucial role in 
transitions between low- and high-confinement states in confined 
plasmas, and their unfoldings have been analyzed.

To our knowledge, TCBs have not been discussed so far for periodic 
orbits in autonomous Hamiltonian systems. In this paper we report
on the occurrence of such bifurcations in a class of two-dimensional 
non-integrable Hamiltonian systems. Since the TCB does not belong 
to the generic bifurcations in two-dimensional symplectic 
maps \cite{meye}, we consider it useful to investigate also the 
mathematical conditions under which it can exist, its stability 
under perturbations of the Hamiltonian, and its unfoldings when 
it is destroyed by a perturbation. For this, we rely on 
mathematical studies by J\"anich \cite{jaen1,jaen2} who introduced 
a class of ``crossing bifurcations'', to which the TCB belongs, 
and derived several theorems and useful formulae for crossing
bifurcations of straight-line librational orbits. Finally, in view 
of the important role which Gutzwiller's semiclassical trace formula 
\cite{gutz} plays for investigations of ``quantum chaos'' (see, e.g., 
\cite{gubu,stoe}), we study the inclusion of transcritically 
bifurcating orbits in the trace formula by an appropriate uniform 
approximation. 

Generic bifurcations of fixed points in two-dimensional symplectic 
maps have been classified by Meyer \cite{meye} in terms of the number 
$m=1,2,\dots$ that corresponds to a period $m$-tupling occurring at 
the bifurcation. For an easily readable presentation of this 
classification of generic bifurcations, and of the corresponding normal
forms used in semiclassical applications, we refer to the textbook of 
Ozorio de Almeida \cite{ozob}. Bifurcations occurring in Hamiltonian 
systems with discrete symmetries have been investigated in 
\cite{rimm,nong,maod,then}; the TCB was, however, not mentioned in
these papers. In \cite{then} it has been shown that all other 
non-generic bifurcations occurring in such systems can be described by 
the generic normal forms given in \cite{ozob}, except for different 
bookkeeping of the number of fixed points which is connected not only
to an $m$-tupling of the period, but also to degeneracies of the 
involved orbits due to the discrete symmetries. For the TCB this is 
not the case: it requires a normal form that is not in the generic 
list of \cite{ozob}. We derive an appropriate normal form 
for the TCB, starting from the general criteria given in \cite{jaen1}, 
and find it to correspond to that given in the literature for
non-Hamiltonian systems \cite{gols,guho}. We use this normal form 
to develop the uniform approximation needed to include transcritically 
bifurcating orbits in the semiclassical trace formula. In a specific 
example that includes a TCB, we show numerically that our result
allows to reproduce the coarse-grained quantum-mechanical density of 
states with a high precision.

In the nonlinear and semiclassical physics community, there exists 
an occasional belief that non-generic bifurcations occur only in 
systems which exhibit discrete symmetries (time-reversal symmetry 
being the most frequently met in physical systems). The examples of
TCBs which we present in this paper are obtained in a class of 
autonomous Hamiltonian systems with mixed dynamics; starting from 
the famous H\'enon-Heiles (HH) Hamiltonian \cite{hh} we change the 
coefficient of one of its cubic terms and add other terms destroying 
some or all of its discrete symmetries. All the TCBs that we have 
found involve one straight-line libration belonging to the shortest 
``period one'' orbits. Our formal investigations therefore focus on the 
class of two-dimensional Hamiltonians containing a straight-line 
librational orbit. In this class of systems the TCB is, in fact, 
found to be the typical isochronous bifurcation of the librating orbit. 
The isochronous pitchfork bifurcation (PFB), however, which in 
Hamiltonian systems with time-reversal symmetry (such as the standard 
HH system) is the most frequently met non-generic bifurcation, is the 
exception here. We show how under a specific perturbation the PFB 
can unfold into a saddle-node bifurcation (SNB) followed by a TCB. 
In a specific example, we demonstrate that the TCB can exist in a 
system without any discrete (spatial or time-reversal) symmetry, thus 
proving that the above-mentioned belief is incorrect. 

Our paper is organized as follows. In \Sec{secmath} we compile
results of J\"anich \cite{jaen1,jaen2} relevant for our investigations. 
Starting from two-dimensional symplectic maps, we define a class of 
``crossing bifurcations'' to which the TCB and the isochronous PFB 
belong. We discuss various criteria and properties of these 
bifurcations and give some useful formulae for the specific case
of a bifurcating straight-line libration. The mathematically less
interested reader may skip \Sec{secmath} and jump directly to
\Sec{secghh}, where we present numerical examples of the TCB and their 
characteristic features in the generalized H\'enon-Heiles Hamiltonians. 
We also study there various types of unfoldings of the TCB under 
perturbations of the Hamiltonian.
In \Sec{secscl} we discuss the semiclassical trace formula for the
density of states of a quantum Hamiltonian, and present the uniform
approximation by which bifurcating periodic orbits can be included.
In \Sec{sectesttf} we present a semiclassical calculation of the
density of states in a situation where the TCB occurs between two 
of the shortest periodic orbits, and demonstrate the validity of the 
uniform approximation by comparison of the results with those of a 
fully quantum-mechanical calculation. In \App{nofosec} we derive the 
appropriate normal forms for the TCB and the isochronous PFB 
which are needed in semiclassical applications. 
In \App{falsetcb} we briefly discuss the stability exchange of two 
orbits in a ``false transcritical bifurcation'' which actually
consists of a pair of close-lying pitchfork bifurcations.

\section{Mathematical prerequisites}
\label{secmath}

In this section we present results of J\"anich \cite{jaen1,jaen2} 
which are relevant for our investigations. We shall only quote 
theorems and other results; for readers interested in the
mathematical proofs or other details, we refer to the explicit
contents of \cite{jaen1,jaen2}.

\subsection{Poincar\'e map and stability matrix}
\label{secappsm}

We are investigating bifurcations of periodic orbits in
two-dimensional Hamiltonian systems. They are most conveniently
investigated and mathematically described by observing the
fixed points on a suitably chosen projected Poincar\'e surface of
section (PSS).\footnote{
With ``projected'' we mean the fact that we ignore the value of the
canonically conjugate variable (e.g., $p_y$), to the variable (e.g., 
$y$) that has been fixed (e.g., by $y=y_0$) to define the true 
mathematical PSS which lies in the energy shell. In the physics 
literature, it is standard to call its projection (with $p_y=0$) the 
PSS. Due to energy conservation, the value of $p_y$ on the unprojected 
PSS can be calculated uniquely, up to its sign which usually is chosen 
to be positive, from the knowledge of $q,p,y_0$ and the energy $E$ 
through the implicit equation $E=H(q,y_0,p,p_y)$, where 
$H(x,y,p_x,p_y)$ is the Hamiltonian in Cartesian coordinates.} 
Since the PSS here is two-dimensional, we describe it by a pair
of canonical variables $(q,p)$. The time evolution of an orbit
then corresponds to the two-dimensional Poincar\'e map
\be
(q,p) \longrightarrow (Q,P)\,,
\label{pcmap}
\ee
where $(q,p)$ is the initial and $(Q,P)$ the final point on the PSS.
Fixed points of this map, defined by $Q=q$, $P=p$, correspond to
periodic orbits.
We introduce $\epsilon$ as a ``bifurcation parameter'' which in
principle may be the conserved energy of the system or any potential
parameter, normalized such that a bifurcation occurs at $\epsilon=0$. 
Here we specialize to the energy variable by defining
\be
\epsilon = E-E_0\,,
\label{eps}
\ee
where $E_0$ is the energy at which the considered bifurcation
takes place. 
We assume that the bifurcating orbit returns to the same point
on the PSS after one map \eq{pcmap}, so that $Q=q$, $P=p$; in 
this paper this will be called a ``period one'' orbit. We shall 
only study its isochronous bifurcations and hence only consider 
the non-iterated Poincar\'e map.

The map \eq{pcmap} is symplectic and thus area conserving in the 
$(q,p)$ plane, and may be understood as a canonical transformation:
\be
Q = Q(q,p,\epsilon)\,, \qquad P = P(q,p,\epsilon)\,.
\label{QPofqp}
\ee
J\"anich \cite{jaen1} has given a classification of bifurcations of 
fixed points in two-dimensional symplectic maps, which we shall 
summarize in the following. We use his notation $Q_u$, $P_u$ for
partial derivatives of the functions $Q$ and $P$, respectively, with 
respect to $u$:
\be
Q_u = \frac{{\partial Q}}{{\partial u}}\,, \qquad 
P_u = \frac{{\partial P}}{{\partial u}}\,,
\label{pardif}
\ee
where $u $ is any of the three variables $q,p$ or $\epsilon$. Analogously 
$Q_{qq}$, $P_{qp\epsilon}$, etc., denote second and higher partial 
derivatives. Due to the symplectic nature of \eq{QPofqp}, the determinant 
of the first derivatives of $Q$ and $P$ with respect to $q$ and $p$ is unity:
\be
{\rm det} \left( \begin{array}{cc} Q_q(q,p,\epsilon) & Q_p(q,p,\epsilon)\\
                                   P_q(q,p,\epsilon) & P_p(q,p,\epsilon)
                 \end{array} \right) = 1\,.
\label{det1}
\ee 

We consider an isolated ``period one'' orbit with fixed point $(q,p)=(0,0)$, 
for all values of $\epsilon$ where it exists, and denote it as the A orbit. 
Its stability matrix is then given by
\be
\MA(\epsilon) = \left( \begin{array}{cc} Q_q(0,0,\epsilon) & Q_p(0,0,\epsilon)\\
                                         P_q(0,0,\epsilon) & P_p(0,0,\epsilon)
                       \end{array} \right).
\label{trma}
\ee
At $\epsilon=0$, where the orbit undergoes an isochronous bifurcation, 
$\MA(0)$ has two degenerate eigenvalues +1, so that $\trMA(0)=2$.

Henceforth we shall omit the arguments (0,0,0) in the partial
derivatives of $Q$ and $P$ which -- unless explicitly mentioned
otherwise -- will always be evaluated at the bifurcation point.
When we need some of these partial derivatives at $p=q=0$ but at
arbitrary values of $\epsilon$, we shall denote them by
$Q_p(\epsilon)$ etc. When no argument is given, $\epsilon=0$ is
assumed. We thus write
\be
\MA(\epsilon) = \left( \begin{array}{cc} Q_q(\epsilon) & Q_p(\epsilon)\\
                                         P_q(\epsilon) & P_p(\epsilon)
                       \end{array} \right),\qquad
\MA(0)        = \left( \begin{array}{cc} Q_q & Q_p\\
                                         P_q & P_p
                           \end{array} \right).
\label{trmaeps}
\ee
The slope of the function $\trMA(\epsilon)$ at $\epsilon=0$ (coming 
from a side where the orbit A exists) becomes, in this notation,
\be
\trMA'(0) = Q_{q\epsilon} + P_{p\epsilon}\,.
\label{slope}
\ee
By a rotation of the canonical coordinates $q,p$ it is always possible
to bring $\MA(0)$ into the form:
\be
\MA(0) = \left( \begin{array}{cc} 1 & Q_p\\
                                  0 & 1
                \end{array} \right), \qquad Q_p\neq0\,.
\label{trma0l}
\ee
We shall henceforth assume that the coordinates have been
chosen such that \eq{trma0l} is true.\footnote{In 
some cases one may find that $\MA(0)$ has the transposed
simple form in which $Q_p=0$ and $P_q\neq 0$. In this
case one may simply exchange the coordinates by a canonical
rotation $Q\rightarrow P$, $P\rightarrow -Q$ (and 
$q\rightarrow p$, $p\rightarrow -q$) in all formulae
below and in \App{whynonf}. The case $Q_p=Q_p=0$ is exceptional 
and occurs only for harmonic potentials (cf.\ \cite{book}).}
Then, with \eq{det1} one finds easily the {\bf determinant 
derivative formula} \cite{jaen1}
\be
Q_{qu}+P_{pu}=Q_p P_{qu} \quad \qquad (u=q,p,\epsilon)\,,
\label{detdera}
\ee
and \eq{slope} takes the simpler form
\be
\trMA'(0) = Q_p P_{q\epsilon}\,.
\label{slopea}
\ee

The total fixed point set
\be
F := \{(q,p,\epsilon)\,|\,Q(q,p,\epsilon)=q,\, P(q,p,\epsilon)=p\}
\label{Fset}
\ee
is the inverse image of the origin $(0,0)$ in $\mathbb{R}^2$ under
the map $(Q-q,P-p)$ whose Jacobian matrix at $(0,0,0)$ is
\be
{\rm J} = \left( \begin{array}{ccc} Q_q-1 & Q_p & Q_\epsilon \\
                                    P_q & P_p-1 & P_\epsilon \\
                 \end{array} \right)
        = \left( \begin{array}{ccc} 0 & Q_p & Q_\epsilon \\
                                    0 & 0   & P_\epsilon \\
                 \end{array} \right).
\label{jacob}
\ee
In the generic case, $P_\epsilon\neq0$ and J has rank 2.
This leads to the only generic isochronous bifurcation
according to Meyer \cite{meye}, the {\bf saddle-node bifurcation} 
(SNB) (also called ``tangent bifurcation''). For this bifurcation, the
fixed-point set $F$ \eq{Fset} is a smooth one-dimensional manifold,
consisting of two half-branches tangent to the $q$
axis at the bifurcation point with slopes $\trMA'(0)=\pm\infty$.
The orbit A then exists either only for $\epsilon\leq0$ or only for
$\epsilon\geq0$; no other orbit takes part in such a bifurcation.

Following J\"anich \cite{jaen1}, we speak of a {\bf rank 1 bifurcation}, 
when the Jacobian J in \eq{jacob} has rank 1, which is the case for
\be
P_\epsilon = 0\,.
\label{rank1}
\ee
Then, after a suitable ($\epsilon$-dependent)
translation of the $p$ variable, J can always be brought into the form
\be
{\rm J} = \left( \begin{array}{ccc} Q_q-1 & Q_p & Q_\epsilon \\
                                    P_q & P_p-1 & P_\epsilon \\
                 \end{array} \right)
        = \left( \begin{array}{ccc} 0 & Q_p & 0 \\
                                    0 & 0   & 0 \\
                 \end{array} \right).
\label{jacob0}
\ee
We shall formulate all following developments in the {\bf suitably 
adapted coordinates} $(q,p)$, for which the form \eq{jacob0} holds, 
and discuss only rank 1 bifurcations.

\subsection{Crossing bifurcations of isolated periodic orbits}
\label{seccross}

A rank 1 bifurcation for which the Hessian
\be
{\rm K} = \left( \begin{array}{cc} P_{qq}        & P_{q\epsilon}\\
                                   P_{q\epsilon} & P_{\epsilon\epsilon}
                 \end{array} \right)
\label{Pqehess}
\ee
at (0,0,0) is regular and {\it indefinite}, i.e., for which 
det\,K $=P_{qq}P_{\epsilon\epsilon}-P_{q\epsilon}^2<0$, shall be 
called a {\bf crossing bifurcation}. J\"anich showed \cite{jaen1} 
that a necessary and sufficient criterion for an orbit A to 
undergo a crossing bifurcation at $\epsilon=0$ is for the slope 
$\trMA'(\epsilon=0)$ to be finite and nonzero. With \eq{trma0l} and
\eq{slopea} we see that
\be
P_{q\epsilon} \neq 0\,
\label{Pqenot0}
\ee
for crossing bifurcations. It follows that if the orbit A undergoes 
a crossing bifurcation at $\epsilon=0$, it exists on both sides of a 
finite two-sided neighborhood of $\epsilon=0$. J\"anich also showed
that for such a bifurcation, the total fixed-point set $F$ \eq{Fset}
is the union $A\cup B$ of two smooth 1-dimensional submanifolds 
intersecting at the bifurcation point. The set $A$ is the set
of fixed points $(0,0,\epsilon)$ of the A orbit; we shall 
call it the {\bf fixed-point branch} $A$. The set $B$ is the fixed-point
set of a second orbit B which takes part in the crossing bifurcation.
 
We shall discuss here only two types of crossing bifurcations:
transcritical and fork-like bifurcations. Their properties are 
specified in the following two subsections. A rank 1 bifurcation
with a regular and {\it definite} Hessian K, i.e., with det\,K $>0$,
is sometimes called an ``isola center'' (cf.\ the normal form
for the isola center in one-dimensional Hamiltonians at the end 
of \Sec{whynonf}). Here the total fixed-point set F consists of the
single isolated point $(q,p,\epsilon)=(0,0,0)$.

\newpage

\subsubsection{Transcritical bifurcation (TCB)} 
\label{sectcb}

A transcritical bifurcation (TCB) occurs when, in the adapted
coordinates $(q,p)$ for which \eq{trma0l} holds, one has
\be
P_{qq} \neq 0\,.
\label{extcl}
\ee
Then, there exists another isolated periodic orbit B on both sides of 
$\epsilon=0$, forming a {\bf fixed-point branch} $B$ intersecting that 
of the orbit A at $\epsilon=0$ with a finite angle. The functions 
$\trMA(\epsilon)$ and $\trM_{\rm B}(\epsilon)$ have opposite slopes 
at the bifurcation:
\be
\trMA'(0)=-\,\trM'_{\rm B}(0)\,. \qquad (\hbox{\it ``TCB~slope~theorem''})
\label{slopestc}
\ee 

In the scenario of a TCB, the orbits A and B simply exchange 
their stabilities and no new orbit appears (or no old orbit 
disappears) at the bifurcation.

{\bf Note:} Assume that the orbit A is a straight-line libration, 
chosen to lie on the $y$ axis, so that the Poincar\'e variables
are $q=x$, $p=p_x$ (see \Sec{seclib} below). Then, if the system 
is invariant under reflexion at the $y$ axis, such a reflexion leads 
to $P(q,p,\epsilon)=-P(-q,-p,\epsilon)$. Therefore, 
$P_{qq}(q=0,p=0,\epsilon=0)=P_{qq}=0$, and the bifurcation 
cannot be transcritical. The simplest possible crossing bifurcation 
then is fork-like (see next item). In short: {\bf Straight-line 
librations along symmetry axes cannot undergo transcritical 
bifurcations.}
 
\subsubsection{Fork-like bifurcation (FLB)} 
\label{secpfb}

A fork-like bifurcation (FLB) occurs when one has
\be
P_{qq}=0\,, \qquad P_{qqq} \neq 0\,.
\label{exfb}
\ee
Then, there exists another isolated periodic orbit B, either 
only for $\epsilon\geq 0$ or only for $\epsilon\leq 0$. The 
fixed-point set of B consists of two half-branches intersecting 
the set $A$ at $\epsilon=0$ at a right angle. In the adapted 
coordinates corresponding to \eq{trma0l}, one may parameterize 
the set $B$ by $(q,p_B(q),\epsilon_B(q))$ and finds
\be
p'_B(0)=\epsilon'_B(0)=0\,, \qquad \epsilon''_B(0)\neq 0\,.
\label{Bbranch}
\ee
Although $\trM_{\rm B}(\epsilon)$ is not a proper function of
$\epsilon$, a limiting slope $\trM'_{\rm B}(0)\neq 0$ can be defined 
for both half-branches of the set $B$ in the limit $\epsilon\to0$, coming 
from that side where they exist, and be shown \cite{jaen1} to fulfill 
the relation 
\be
\trM'_{\rm B}(0)=-2\,\trMA'(0)\,. \qquad (\hbox{\it ``FLB~slope~theorem''})
\label{slopesfb}
\ee 
In the same limit, the curvature of the set $B$ at the bifurcation 
point is given by
\be
\epsilon''_B(0) = \frac{3Q_{qq}P_{qp}-Q_pP_{qqq}}
                       {3Q_pP_{q\epsilon}}\,.
\label{epspp}
\ee
               
In the pertinent physics literature, this bifurcation is often called 
the (non-generic) {\bf isochronous pitchfork bifurcation (PFB)}. Note 
that here the two half-branches of the set $B$ correspond to two
different periodic orbits. They can be either locally degenerate (to
first order in $\epsilon$), or globally degenerate due to a discrete 
symmetry (reflexion at a symmetry axis or time reversal). 

In the {\bf generic PFB} corresponding to Meyer's classification 
\cite{meye}, the fixed point scenario near $\epsilon=0$ is identical 
with that of the FLB. However, here the two fixed points of the set 
$B$ correspond to one single orbit B which has twice the period of 
the primitive orbit A. In fact, the fixed-point branch $A$ crossing 
the line $\trMA=2$ is that of the {\bf iterated} Poincar\'e map: 
the generic PFB is {\bf period doubling}. The existence criterion 
\eq{exfb} and the relations \eq{Bbranch} - \eq{epspp} for the B orbit 
hold here, too \cite{ssu1}.

\newpage

\subsection{Some explicit formulae for straight-line librations}
\label{seclib}

\subsubsection{Definition of the librational A orbit}

We now specialize to straight-line librational orbits in two-dimensional 
autonomous Hamiltonian systems, defined by Hamiltonian functions
\be
H_0(x,y,p_x,p_y) = \frac12\,(p_x^2+p_y^2) + V(x,y)
\label{ham0}
\ee
with a smooth potential $V(x,y)$. Straight-line librations form the 
simplest type (and so far the only one known to us) of periodic orbits 
in Hamiltonian systems that undergo transcritical bifurcations. Let
us choose the direction of the libration to be the $y$ axis and call
it the A orbit. The potential then must have the property
\be
\frac{\partial V}{\partial x}\,(0,y) = 0
\ee
for all $y$ reached by the libration. The A orbit, which we assume 
to be bound at all energies, then has $x(t)={\dot x}(t)=0$ for all 
times $t$, and its $y$ motion is given by the Newton equation
\be
{\ddot y}(t) + \frac{\partial V}{\partial y}\,(0,y(t))=0 \qquad
\Rightarrow \qquad y(t) = y_A(t,\epsilon)\,,
\label{eomya}
\ee
where $y_A(t,\epsilon)$ is henceforth assumed to be a known periodic
function of $t$ with period $T_A(\epsilon)$. For the
A orbits in the (generalized) HH potentials discussed in the following 
section, the function $y_A(t,\epsilon)$ can be expressed in terms of a 
Jacobi-elliptic function \cite{lame}. We choose the time scale such
that $y_A(0,\epsilon)$ is maximum with 
\be
{\dot y}_A(0,\epsilon)=0 \qquad \forall\;\epsilon\,.
\label{slopya}
\ee
A suitable choice of Poincar\'e variables is to use the surface of 
section defined by $y=0$, and the projected PSS becomes the $(x,p_x)$ 
plane, so that we define $q=x$, $p=p_x$. We again assume that the 
orbit A is isolated and exists in a finite interval of $\epsilon$ 
around zero. The fixed-point branch  $A$ is thus again given by the 
straight line $(q_A,p_A,\epsilon)=(0,0,\epsilon)$ in the 
$(q,p,\epsilon)$ space. 

In \cite{jaen2} J\"anich has given an iterative scheme to calculate 
the partial derivatives $Q_q$, $Q_p$, etc.\ for this situation for 
any given (analytical) potential $V(x,y)$ with the above properties. 
To this purpose, one has first to determine the fundamental systems 
of solutions $(\xi_1,\xi_2)$ and $(\eta_1,\eta_2)$ of the linearized 
equations of motion in the $x$ and $y$ directions, respectively:
\bea
{\ddot \xi} (t) + V_{xx}(0,y_A(t,\epsilon))\,\xi(t)
 & = & 0\,,\label{linx}\\
{\ddot \eta}(t) + V_{yy}(0,y_A(t,\epsilon))\,\eta(t)
 & = & 0\,,\label{liny} 
\eea
with the initial conditions 
\be
\left(\begin{array}{cc} \xi_1(0) & \xi_2(0) \\
                       {\dot\xi}_1(0) & {\dot\xi}_2(0) \end{array}\right) =
\left(\begin{array}{cc} \eta_1(0) & \eta_2(0) \\
                       {\dot\eta}_1(0) & {\dot\eta}_2(0) \end{array}\right)=
\left(\begin{array}{cc} 1 & 0 \\
                        0 & 1 \end{array}\right)\qquad \forall\;\epsilon.
\label{initcond}
\ee
For simplicity, we do not give the argument $\epsilon$ of the $\xi_i(t)$ 
and $\eta_i(t)$, but we should keep in mind that they are all functions 
of $\epsilon$. In \eq{linx}, \eq{liny} the subscripts on the function 
$V(x,y)$ denote its second partial derivatives with respect to the 
corresponding coordinates. In the formulae given below, we denote by
$V_i(t)$, $V_{ij}(t)$, etc., with $i,j\in(x,y)$ the partial derivatives 
taken along the A orbit, i.e., at $x=0$, $y=y_A(t,\epsilon)$ as in
\eq{linx}, \eq{liny}, evaluated at the bifurcation point $\epsilon=0$. 
If the partial derivatives have no argument, they are taken at the 
period $T_A(\epsilon_0)$, i.e., $V_y=V_y(T_A(\epsilon_0))$ etc.

Knowing the five functions $y_A(t,\epsilon=0)$ and $\xi_i(t),\eta_i(t)$
$(i=1,2)$ at $\epsilon=0$, all desired partial derivatives of 
$Q(q,p,\epsilon)$ and $P(q,p,\epsilon)$ at $(q,p,\epsilon)=(0,0,0)$ can be 
obtained by (progressively repeated) quadratures, i.e., by finite integrals 
over known expressions including these five functions, partial derivatives 
of $V(x,y)$, and the functions obtained at earlier steps of the scheme
(whereby the progression comes from increasing degrees of the desired
partial derivatives). 

\subsubsection{Stability matrix of the A orbit}

We note that the equation \eq{linx} is nothing but the stability 
equation of the A orbit, since the $\xi_i$ by definition are small 
variations transverse to the orbit. In the standard literature, 
\eq{linx} is also called the ``Hill equation'' (cf., e.g., 
\cite{gubu,mawi}). The stability matrix $\MA$ at the bifurcation 
of the A orbit is therefore simply given by
\be
\MA(0) = \left( \begin{array}{cc} Q_q & Q_p\\
                                  P_q & P_p
                \end{array} \right)
       = \left( \begin{array}{cc}      \xi_1(T_A)  &       \xi_2(T_A)\\
                                  {\dot\xi}_1(T_A) & {\dot\xi}_2(T_A)
                \end{array} \right),
\label{trma0sol}
\ee
with $T_A=T_A(\epsilon=0)$. Its eigenvalues must be $\lambda_1=
\lambda_2=+1$, as seen directly from \eq{trma0l}. The solutions 
$\xi_i(t,\epsilon)$ of \eq{linx} are in general not periodic. 
But at the bifurcations of the A orbit, where $\trM_A=+2$, one 
of the $\xi_i(t,\epsilon=0)$ is always periodic with period $T_A$ 
(or an integer multiple $m$ thereof) \cite{mawi} and describes, up 
to a normalization constant depending on $\epsilon$, the transverse 
motion of the bifurcated orbit at an infinitesimal distance 
$\epsilon$ from the bifurcation (cf.\ \cite{lame,mbgu,bfmm}).


\subsubsection{Slope of the function $\trMA(\epsilon)$ at $\epsilon=0$}
\label{secslopa}

Here we give the explicit formulae, obtained from \cite{jaen2},
for the slope $\trMA'(0) = Q_{q\epsilon} + P_{p\epsilon}$, see 
\eq{slope}, of the function $\trMA(\epsilon)$ at the bifurcation.
The quantities $Q_{q\epsilon}$ and $P_{p\epsilon}$ are given, in
terms of the potential $V(x,y)$ in \eq{ham0} and the other
ingredients defined above, by
\bea
Q_{q\epsilon} & = & \;\frac{1}{(V_y)^2}\,P_q\,{\dot\eta}_1(T_A)
                  - \frac{1}{V_y}\int_0^{T_A} V_{xxy}(t)
                    \left[Q_p\,\xi_1(t)-Q_q\,\xi_2(t)\right]
                    \xi_1(t)\,\eta_1(t)\,{\rm d}t\,,\label{Qqeps}\\
P_{p\epsilon} & = & \frac{(-V_{xx})}{(V_y)^2}\,Q_p\,{\dot\eta}_1(T_A)
                  - \frac{1}{V_y}\int_0^{T_A}\! V_{xxy}(t)
                    \left[P_p\,\xi_1(t)-P_q\,\xi_2(t)\right]
                    \xi_2(t)\,\eta_1(t)\,{\rm d}t\,.\label{Ppeps}
\eea
In the adapted coordinates where $\trMA(0)$ has the form \eq{trma0l} with
$Q_q=P_p=1$ and $P_q=0$, the slope becomes
\be
\trMA'(0) = Q_p\,P_{q\epsilon}
          = \frac{(-V_{xx})}{(V_y)^2}\,Q_p\,{\dot\eta}_1(T_A)
            - \frac{1}{V_y}\,Q_p \int_0^{T_A} V_{xxy}(t)\,
              \xi_1^2(t)\,\eta_1(t)\,{\rm d}t\,.
\label{slopot}
\ee
For the case that $\trMA(0)$ has the transposed tridiagonal form
with $Q_p=0$ and $P_q\neq0$, the formula becomes
\be
\trMA'(0) = P_q\,Q_{p\epsilon}
          = \frac{1}{(V_y)^2}\,P_q\,{\dot\eta}_1(T_A)
            + \frac{1}{V_y}\,P_q \int_0^{T_A} V_{xxy}(t)\,
              \xi_2^2(t)\,\eta_1(t)\,{\rm d}t\,.
\label{slopotr}
\ee

An independent derivation of \eq{Qqeps} - \eq{slopotr} is given in
\cite{serg}, where it is shown that the first terms are due to the
variation of the A orbit's period $T_A$ with $\epsilon$, whereas
the integral terms are due to the $\epsilon$ dependence of the
functions $\xi_i(t)$. 

\subsubsection{Criterion for the TCB}

For a bifurcation to be transcritical, we need $P_{qq}\neq 0$.
From \cite{jaen2} we find the following explicit formula for
$P_{qq}$
\be
P_{qq} = -\int_0^{T_A} V_{xxx}(t)\,\xi_1^3(t)\,{\rm d} t\,,
\label{Pqq}
\ee
which also yields explicitly the parameter $b$ in its normal
form given in \eq{nftcb} below.

If the potential is symmetric about the $y$ axis, then $V_{xxx}(t)$ 
is identically zero and the TCB cannot occur, as already stated in 
\Sec{sectcb} above. However, even if $V_{xxx}(t)$ is not zero,
special symmetries of the function $\xi_1(t)$, in combination with 
that of $V_{xxx}(t)$, can make the integral in \eq{Pqq} vanish. An
example of this is discussed in \Sec{secunfpfb}.

\newpage

\section{TCBs in the generalized H\'enon-Heiles potential}
\label{secghh}

\subsection{The generalized H\'enon-Heiles potential}

For our numerical studies, we have investigated the following family
of generalized H\'enon-Heiles (GHH) Hamiltonians:
\be
H(x,y,p_x,p_y) = \frac12\, (p_x^2+p_y^2) + \frac12\,(x^2+y^2)
                 + \alpha\left[-\frac13\,y^3 + \gamma\,x^2y+\beta\,y^2x\right].
\label{ghh}
\ee
Here $\alpha$ is the control parameter that regulates the nonlinearity
of the system, and $\gamma$, $\beta$ are parameters that define
various members of the family.
The standard HH potential \cite{hh} corresponds to $\gamma=1$, $\beta=0$.
It has three types of discrete symmetries: ($i$) rotations about
$2\pi/3$ and $4\pi/3$, and ($ii$) reflections at three corresponding 
symmetry lines, which together define the C$_{3v}$ symmetry, and
($iii$) time-reversal symmetry.
There exist three saddles at the critical energy $E^*=1/6\alpha^2$, so
that the system is unbound and a particle can escape if its energy is 
$E>E^*$. For $\gamma\neq 1$, $\beta\neq 0$, the spatial symmetries are 
in general broken (except for particular values of $\gamma$ and
$\beta$) and only the time-reversal symmetry is left. There still
exist three saddles, but in general they lie at different energies. 
There is always a stable minimum at $x=y=0$.

It is convenient to scale away the nonlinearity parameter $\alpha$ in 
\eq{ghh} by introducing scaled variables $x'=\alpha x$, $y'=\alpha y$ 
and a scaled energy $e=E/E^*=6\alpha^2E$. Then \eq{ghh} becomes
\be
h = e = 6\alpha^2 E = 3\,(p_{x'}^2 + p_{y'}^2+x'^2+y'^2)
                  - 2\,y'^3 + 6\,(\gamma\, x'^2y'+\beta\, y'^2x')\,,
\label{ghhsc}
\ee
so that one has to vary one parameter less to discuss the classical
dynamics. (For the standard HH potential with $\gamma=1$, $\beta=0$, 
the scaled energy $e$ then is the only parameter.) For simplicity, we 
omit henceforth the primes of the scaled variables $x',y'$.

Before we discuss the periodic orbits in the system \eq{ghhsc}, let us 
briefly recall the situation in the standard HH system in which all
three saddles lie at the scaled energy $e=1$.

\subsubsection{Periodic orbits in the standard HH potential}
\label{secshh}

The periodic orbits of the standard HH system have been studied in 
\cite{lame,mbgu,chup,hhpo}, and their use in connection with semiclassical 
trace formulae in \cite{hhsc,hh1,pert,hhun,kaid,kwb}. We also refer to 
\cite{book} (section 5.6.4) for a short introduction into this system,
which represents a paradigm of a mixed Hamiltonian system covering
the transition from integrability ($e=0$) to near-chaos ($e>1$). 

In \fig{hhtr} we show the trace tr\,M of the stability matrix M,
henceforth called ``stability trace'', of the shortest orbits as a
function of $e$. Up to energy $e\simeq 0.97$, there exist \cite{chup}
only three types of ``period one'' orbits [in the sense defined after 
\eq{eps}]: 1) straight-line librations A
along the three symmetry axes, oscillating towards the saddles; 2)
curved librations B which intersect the symmetry lines at right angles 
and are hyperbolically unstable at all energies; and 3) rotational 
orbits C in the two time-reversed versions which are stable up to 
$e\simeq 0.89$ and then become inverse-hyperbolically unstable. While 
the B and C orbits exist at all energies, the orbits A cease to exist 
at the critical saddle energy $e=1$ where their period becomes infinite.

When $|$tr\,M$| >$ 2 or $<$ 2, an orbit is unstable or stable, respectively.
When tr\,M = 2 it either undergoes a bifurcation if the orbit is isolated,
or it belongs to a family of degenerate orbits in the presence of a continuous
symmetry. The latter is seen to occur in the limit $e\rightarrow 0$, where
the orbits A, B and C all converge to the family of orbits of the isotropic
two-dimensional harmonic oscillator with U(2) symmetry. The A orbits undergo 
an infinite sequence of (non-generic) isochronous PFBs, starting at
$e\simeq 0.97$ and cumulating at $e=1$. At these bifurcations an
alternating sequence of rotational orbits (labeled R) and librational
orbits (labeled L) are born. This bifurcation cascade, the R and L
type orbits, and their self-similarity have been discussed extensively
in \cite{lame,mbgu}. In \fig{hhtr} and in the text below, we indicate 
their Maslov indices (needed in semiclassical trace formulae, see 
\Sec{secscl}) by
suffixes to their labels, which allows for unique bookkeeping of all 
orbits. (Only the first two representatives R$_5$ and L$_6$ of the 
orbits born along the bifurcation cascade are shown in \fig{hhtr} by
the dashed lines.) At each bifurcation, the orbit A increases its
Maslov index (which is 5 up to the first bifurcation) by one unit. 
Only the first three bifurcations can be seen in the figure; the
others are all compressed into a tiny interval below $e=1$. As has
been observed numerically in \cite{lame,mbgu,hhpo}, tr\,M$_A$ becomes 
a periodic function of the period $T_A$ in the limit $e\rightarrow 1$. 
[It can actually be rigorously shown that, asymptotically, tr\,M$_A(T_A)
\longrightarrow -2.68042\,\sin(\sqrt{3}\,T_A)$ in this limit 
\cite{serg}.]

\Figurebb{hhtr}{50}{35}{995}{470}{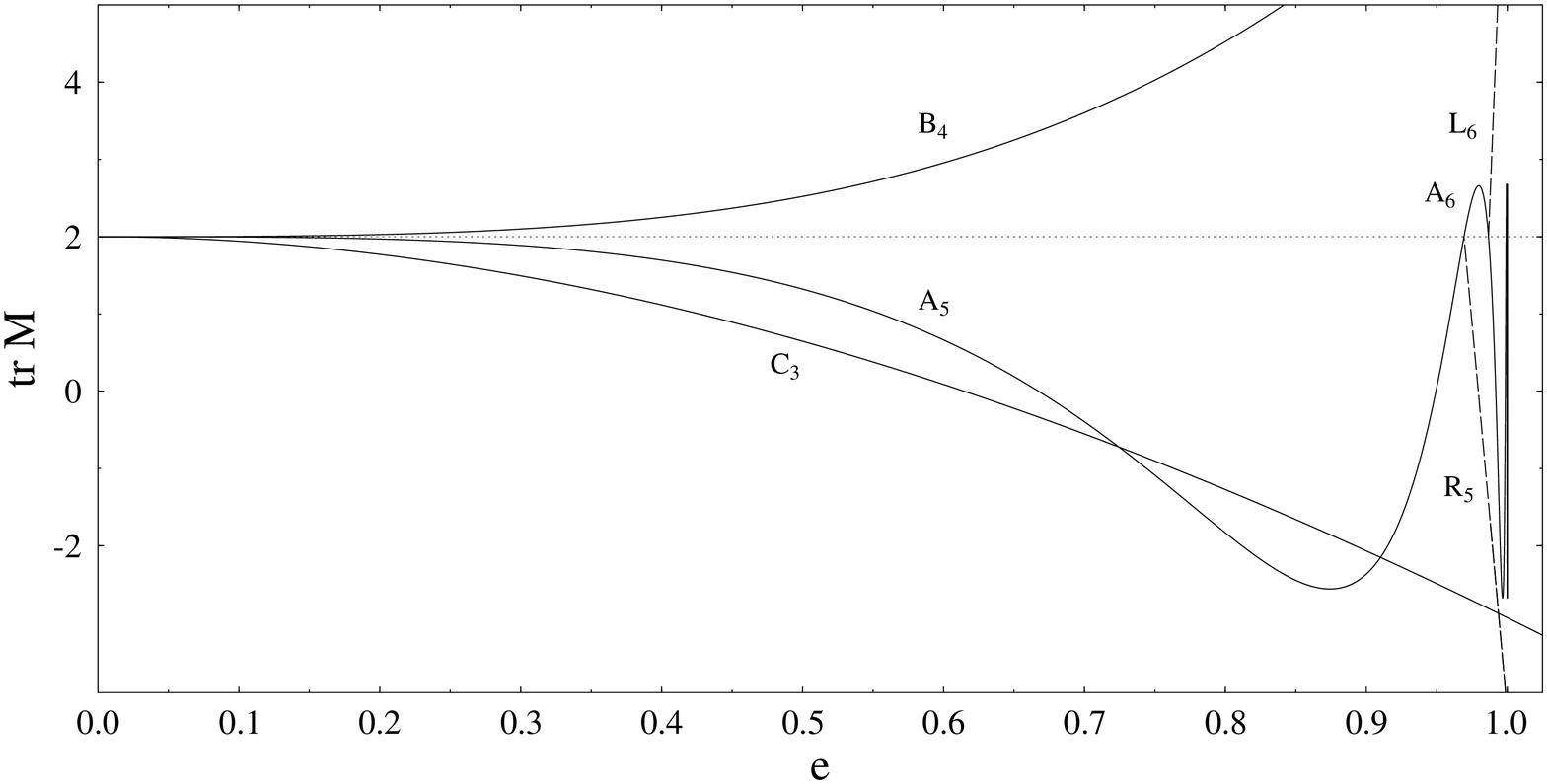}{7.5}{16.5}{
Trace of the stability matrix M of the ``period one'' orbits 
in the standard HH potential, plotted versus scaled energy $e$. 
The suffixes indicate their Maslov indices. Only the first
two (R$_5$ and L$_6$) of the orbits born at an infinite
sequence of isochronous PFBs of the A orbit, cumulating at the 
saddle energy $e=1$, are shown (dashed lines).
}
\vspace*{-0.5cm}

As is characteristic of isochronous PFBs (cf. \Sec{secpfb}), the
new-born orbits come in degenerate pairs due to the discrete symmetries: 
the two librational L orbits are mapped onto each other under 
reflection at the axis containing the A orbit, and the two rotational 
R orbits are connected by time reversal. Note that although the
A orbit ceases to exist for $e\geq1$, all R and L type orbits 
bifurcated from it exist at all energies $e\geq1$. For some new 
orbits appearing there, we refer to the literature \cite{hhpo,kwb};
in this paper we shall not be concerned with them.

\subsubsection{Periodic orbits in the generalized HH potential}

For $\gamma\neq 0$, $\beta\neq 0$, there exist in general three different 
saddles at scaled energies $e_0$, $e_1$ and $e_2$, and three different 
straight-line periodic orbits, labeled A, A' and A'', oscillating towards 
the saddles. In general, there are three curved librational orbits B, B'
and B'' (not necessarily existing at all energies) intersecting the
three A type orbits at right angles, and
there is always a time-reversally degenerate pair of rotational orbits
C going around the origin. It is rather easy to see that the three A 
type orbits always intersect each other at the minimum of the potential
located at the origin $(x,y)=(0,0)$. The equations of motion 
for the Hamiltonian \eq{ghh} in the Newton form are (in the scaled
variables corresponding to $\alpha=1$)

\newpage

\bea
\ddot{x} + x\,(1+2\gamma\,y)+\beta\,y^2 & = & 0\,,
               \nonumber\\
\ddot{y} + y\,(1+2\beta\,x)+\gamma\,x^2-y^2 & = & 0\,. 
\label{eom}
\eea
For a straight-line orbit librating through the origin we have
$y = ax$ which, inserted into \eq{eom}, yields a cubic equation for 
the slope $a$:
\be
\beta\,a^3+(2\gamma+1)\,a^2-2\beta\,a-\gamma = 0\,.
\label{slopeq}
\ee
For the rest of this paper, we limit the parameters to the range
$\gamma>0$ and $\beta\geq 0$. Then, \eq{slopeq} has always real 
roots that are in general different. In the right panel of \fig{trace}
below, we have shown the six shortest (``period one'') librations 
obtained numerically for $\gamma=0.6$, $\beta=0.07$, including the 
three straight-line orbits A, A', A'' intersecting at the origin.

Further analytical analysis is cumbersome except for the following 
special cases:

\noindent
{\it $\beta=0$, $\gamma=1$ (standard HH)}\,:\\
Two of the slopes are $a_{1,2}=\pm 1/\!\sqrt{3}$; the third is 
$a_0=\infty$ corresponding to the orbit along the $y$ axis with 
$x(t)=0$. The three saddles lie at $(x,y)=(0,1)$, $(-\sqrt{3}/2,-1/2)$, 
and $(\sqrt{3}/2,-1/2)$, forming an equilateral triangle with side
length $\sqrt{3}\,$; its sides (and their extensions) form the 
equipotential lines for $e=1$. The periodic orbits are those discussed
in \Sec{secshh}.

\noindent
{\it $\beta=0$, $\gamma\neq 1$}:\\
The rotational $C_{3v}$ symmetry is broken, but the reflection symmetry at 
the $y$ axis is kept. Correspondingly, we find two degenerate orbits 
A', A'' with opposite slopes $a_{1,2}=\pm\sqrt{\gamma/(2\gamma+1)}$.
There is a horizontal equipotential line at $y_1=y_2=-1/2\gamma$ with 
scaled energy $e_1=e_2=(3+1/\gamma)/4\gamma^2$ that contains two saddle 
points symmetrically positioned at $x_{1,2}=\pm\sqrt{(2+1/\gamma)}/2\gamma$.
At low energies, there is only one B type orbit intersecting 
the $y$ axis at a right angle; two further orbits B' and B'' appear through
bifurcations at higher energies (see examples in \Sec{sectra}). For
$\gamma>0$ there is a third A orbit librating along the $y$ axis 
($a_0=\infty$) towards a third saddle at $(0,1)$ with energy $e_0=1$. 
The equipotential line for $e=e_{1,2}$ consists of the 
horizontal line at $y_{1,2}=-1/2\gamma$ and two branches of a hyperbola. 
For $\gamma>1$ the hyperbola branches lie symmetrically about the $y$ axis, 
each intersecting the horizontal line at one of the two symmetric saddle 
points. For $0<\gamma<1$, they lie symmetrically about a horizontal line
at $y^*=(1+3\gamma)/4\gamma$, the lower of them intersecting the line 
$y=y_{1,2}$ at the two symmetric saddle points.

The limiting case $\beta=0$, $\gamma=0$ yields a separable and 
hence integrable system with only one saddle
at $(0,1)$ at energy $e_0=1$ and one A orbit (with $a=\infty$). We do 
not discuss this system here, but refer to \cite{kaid} in which it is 
investigated both classically and semiclassically in full detail.


\subsection{Examples of transcritical bifurcations and their properties}
\label{sectra}

As mentioned above, we have restricted the parameters $\gamma$ and 
$\beta$ in the GHH potential \eq{ghh} to be positive (or $\beta=0$).
We find that, depending on the values of $\beta$ and $\gamma$, at
least one or two of the straight-line orbits A, A', or A'' can 
undergo a TCB with a partner of the curved librational orbits B, B', 
or B''. In the following, we shall first show two examples and then 
discuss characteristic properties of the TCB. In \Sec{secunftcb} we 
shall study its stability and its unfoldings. Some of the numerical
results can easily be understood analytically in terms of the normal 
forms of the various bifurcations and their unfoldings. These are 
discussed in detail in \App{nofosec} and shall be referred to in the 
following text.

\subsubsection{Two examples}

As a numerical example, we choose $\gamma=0.6$, $\beta=0.07$. The
three saddle energies are e$_0=0.993$ for the A orbit, $e_1=2.81$ 
for the A' orbit, and $e_3=3.74$ for the A'' orbit. In the left panel
of \fig{trace} we show the stability traces tr\,M($e$) of the 
shortest orbits. In the right panel we display the shapes of these 
orbits in the $(x,y)$ plane. The orbits B' and B'' are created in a 
SNB at $e_{tb}\simeq 1.533$ and do not exist
below this energy; at high energies they are hyperbolically unstable 
with increasing Lyapunov exponents. Contrary to the standard HH system,
only the A orbit is stable at low energies, while the orbits A' and A'' leave
the $e=0$ limit unstable and cross the critical line tr\,M = +2 at some
finite energies $e_{tcb}$ and $e'_{tcb}$ to become stable. At higher 
energies, all three A type orbits undergo an infinite PFB cascade
as in \fig{hhtr}, each of them converging at its saddle energy. 
(We do not show here the R and L type orbits born at these 
bifurcations.) 

It is between the pairs of orbits A', B and A'', B' that we here observe 
two TCBs. They occur at the energy $e_{tcb}=0.854447$ between the 
orbits A' and B, and at $e'_{tcb}=1.644$ between the orbits A'' and 
B'. The situation near $e'_{tcb}$ actually displays an example of a 
slightly broken PFB which will be discussed in \Sec{secunfpfb}.

\vspace*{-0.5cm}
\Figurebb{trace}{75}{30}{1300}{511}{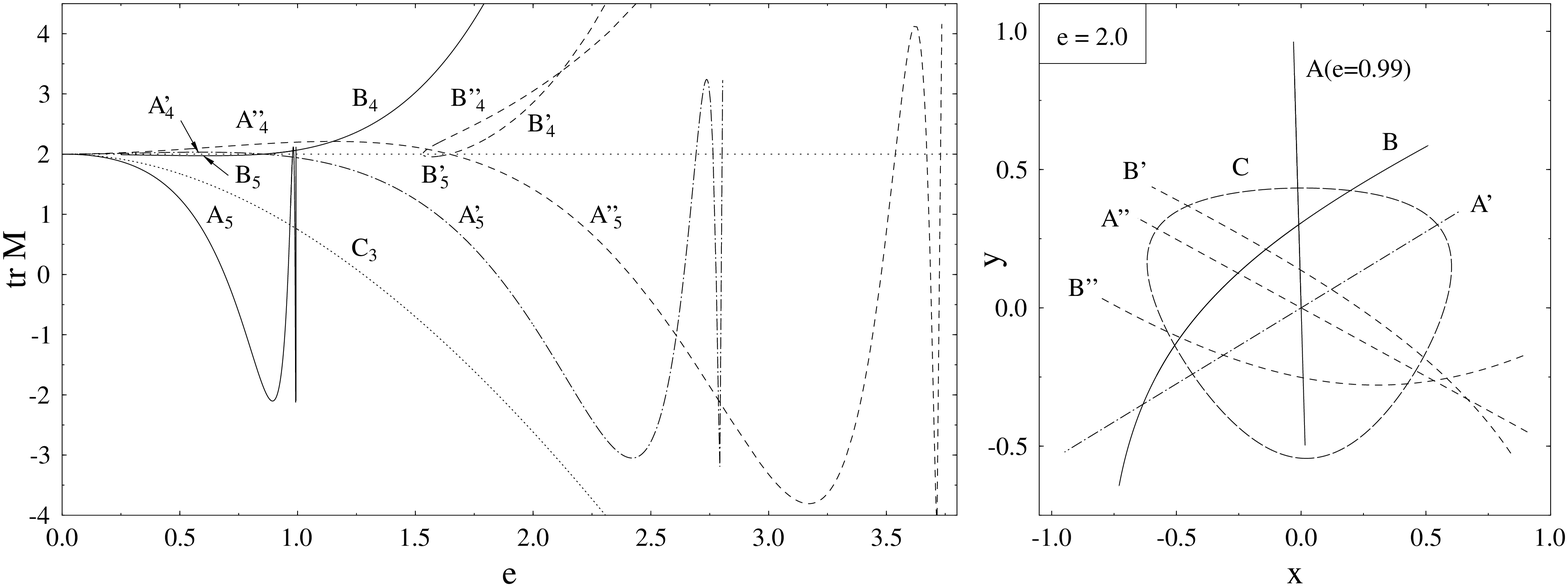}{6.6}{16.5}{
{\it Left panel:}
Stability traces of A and B type orbits in the GHH potential 
with $\gamma=0.6$, $\beta=0.07$, plotted versus scaled energy $e$. 
The three saddles are at $e_0=0.993$, $e_1=2.81$, and $e_2=3.74$.
{\it Right panel:}
Shortest orbits, projected on the $(x,y)$ plane. Orbit A is evaluated 
at $e=0.99$ just below its saddle ($e_0=0.993$); all other orbits are 
taken at $e=2$. The line types correspond to those in the left panel.
}
\vspace*{-0.5cm}

\Figurebb{tcbdeg}{20}{50}{709}{250}{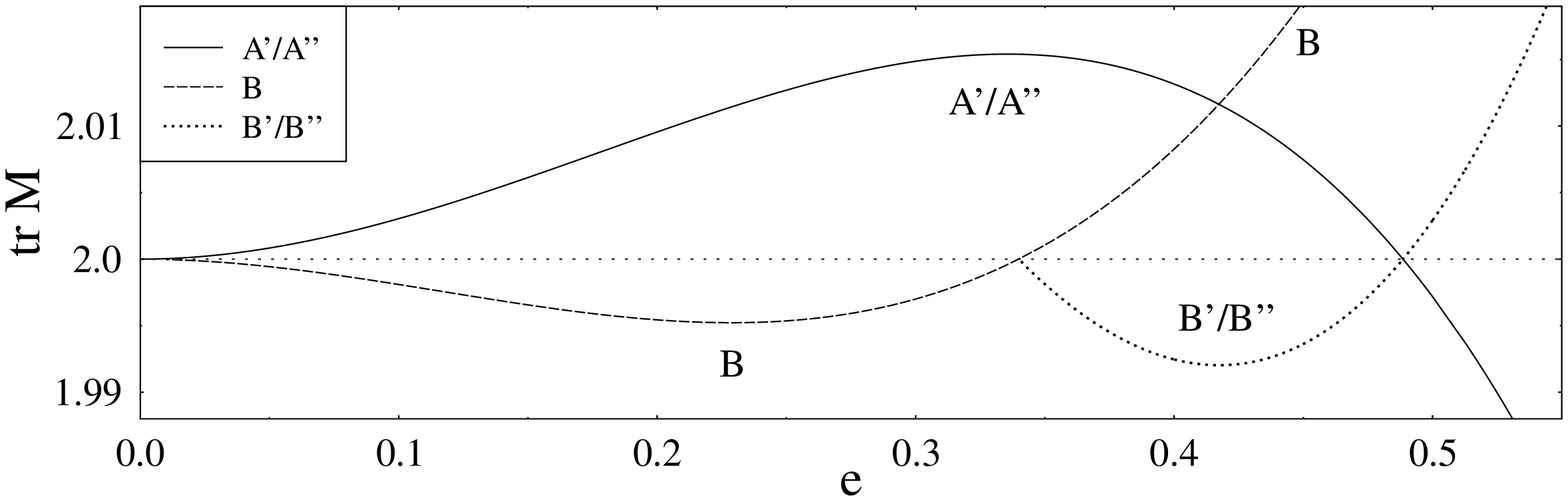}{3.7}{16.5}{
TCB of a degenerate pair of orbits A', A'' and B', B'' at
$e_{tcb}=0.4889$ in the GHH potential with $\gamma=0.75$, 
$\beta=0$. The degeneracy is due to the reflection symmetry about
the $y$ axis.
}
\vspace*{-0.5cm}

Another example of a TCB is shown in \fig{tcbdeg}, obtained
for the GHH potential with  $\gamma=0.75$ and $\beta=0$.
This potential is symmetric about the $y$ axis and
therefore the pairs of orbits A', A'' and B', B'' are degenerate,
lying opposite to each other with respect to the $y$ axis.
The crossing happens at $e_{tcb}=0.4889$ and exhibits the same
features as those discussed in the first example.


\subsubsection{Characteristic properties of the TCB}

We now discuss some of the properties of a TCB and compare our 
numerical results to their analytical predictions from the normal
form of the TCB. For this purpose, we take the example at 
$e_{tcb}=0.854447$ seen in \fig{trace}, where the orbits A' and B 
bifurcate transcritically. Their crossing is shown in \fig{cross0} 
on an enlarged scale in the upper left panel, where the numerical 
results for tr\,M$(e)$ are displayed by crosses (orbit A') and 
circles (orbit B). We see that the graphs of tr\,M($e$) cross the
critical line tr\,M = 2 with opposite slopes. Their Maslov indices,
differing by one unit, are exchanged at the bifurcation 
(see Secs.~\ref{secscl} and \ref{secunifo}). The upper right panel 
displays the numerical action difference $\Delta S=S_B-S_{A'}$ 
(circles), where the action of each periodic orbit (po) is, as 
usual, given by
\be
S_{po} = \oint {\bf p}\cdot {\rm d}{\bf q}\,.
\label{action}
\ee
In the lower panels, we show the shapes of the orbits in the 
$(x,y)$ plane below (left) and above (right) the TCB. The B orbit 
is seen to have passed through the A' orbit at the bifurcation.
The lengths of both orbits increase with energy $e$.
\Figurebb{cross0}{10}{10}{767}{480}{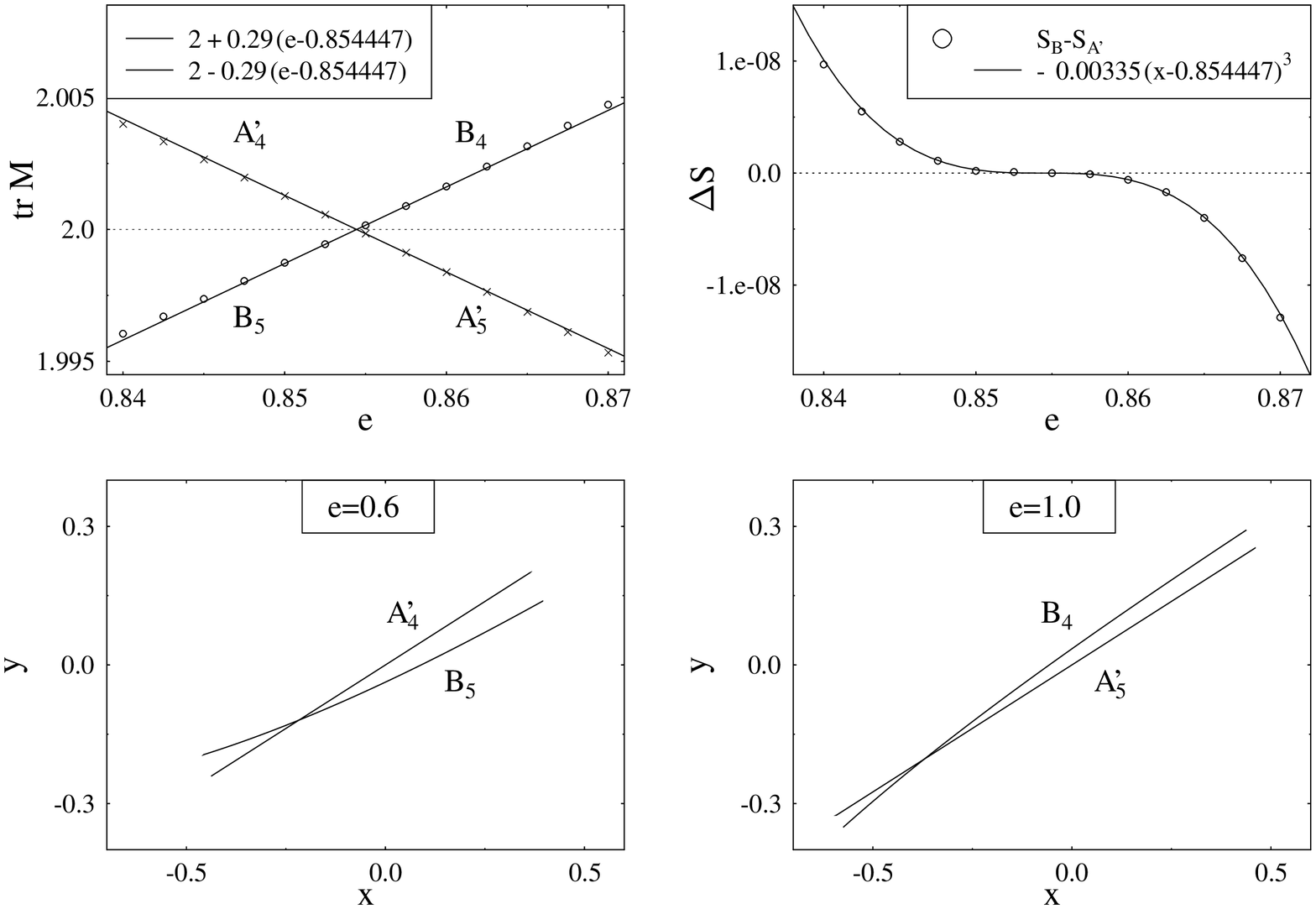}{9.8}{16.5}{
TCB in the GHH potential with $\gamma=0.6$, 
$\beta=0.07$. Orbits A' and B exchange their stabilities 
(and Maslov indices $\sigma=$ 4,5) at $e_{tcb}=0.854447$.
{\it Upper left:} tr\,M versus energy $e$; 
crosses (A') and circles (B) are numerical results, solid lines the 
local prediction \eq{trMn}.
{\it Upper right:} action difference $\Delta S$ versus $e$; circles 
are numerical results
and the solid line the local prediction \eq{delSntcb}. {\it Lower
panels:} shapes of the crossing orbits in the $(x,y)$ plane before
and after the bifurcation.}
\vspace*{-0.6cm}

The normal form of the TCB is derived and discussed in \Sec{secnormtcb}. 
From it, one can derive the local behavior of the actions, periods, 
and stability traces of the two orbits in the neighborhood of a TCB. 
For small deviations $\epsilon=c\,(e-e_{tcb})$ (with $c>0$) from the 
bifurcation energy, the stability traces go like tr\,M$(\epsilon)=
2\pm2\sigma\epsilon$, and the action difference of the two orbits like 
$\Delta S(\epsilon)=-\epsilon^3\!/6b^2$ (see \App{secnormtcb} for the 
meaning of the other parameters). These local predictions, given in 
\fig{cross0} by the solid lines, can be seen to be well followed by 
the numerical results.


The crossing of the graphs tr\,M($e$) of the two orbits at the 
bifurcation energy $e_{tcb}$ with opposite slopes is a characteristic 
feature of the TCB (see \Sec{sectcb}). Since the fixed points of the 
two orbits coincide at the bifurcation point, their shapes must be 
identical there. In the present example, the orbit B is a curved 
libration; the sign of its curvature is changed at the bifurcation, 
as illustrated in the two lower panels of \fig{cross0}.

We note that a completely different mechanism of stability exchange 
of two orbits, which happens through two close-lying PFBs, has been 
described in \cite{erda}. The stability diagram may then appear like 
that in the upper left of \fig{cross0}, if the crossing point is not 
analyzed with sufficient numerical resolution. Such a ``false
transcritical bifurcation'' will be briefly discussed and illustrated 
in the appendix of our paper.


\subsection{Stability and unfoldings of the TCB}
\label{secunftcb}

Since the TCB is not a generic bifurcation according to Meyer's list 
\cite{meye}, we now address the question under which circumstances 
it can exist and what its structural stability is. The GHH systems 
discussed here have time-reversal symmetry, and it is therefore of 
interest to study the stability of the TCB under perturbations of 
the Hamiltonians that destroy this symmetry. In this context, it is 
important to note that a detailed mathematical study \cite{rimm}, 
in which all generic bifurcations in systems with time-reversal 
symmetry are classified, does not mention the TCB; the same holds 
also for \cite{then}. So far we have only found TCBs which involve 
a straight-line libration. On the basis of the results presented in 
\Sec{secmath}, we believe that in the class of all Hamiltonian systems 
containing a straight-line librating orbit, the TCB is actually
the generic isochronous bifurcation of the librating orbit. Therefore, 
if we find a perturbation of the GHH system that destroys the 
time-reversal symmetry but preserves a straight-line libration, the 
TCB should also exist there. This will be demonstrated in 
\Sec{secsurv} for a specific example.

A general Hamiltonian $H(x,y,p_x,p_y)$ supports the existence of
a straight-line libration -- which, without loss of generality, may
be chosen to lie on the $y$ axis -- if the following conditions are 
fulfilled:
\be
\papa{H}{x}\,(0,y,0,p_y) = 0\,, \qquad
\papa{H}{p_x}\,(0,y,0,p_y) = 0\,.
\label{libcond} 
\ee
In the following we will first show how some TCBs are destroyed under 
perturbations that violate the conditions \eq{libcond}, and how they 
unfold. We find two types of unfoldings which are also discussed in 
\cite{bad1,gols} for TCBs in non-Hamiltonian systems. In the first 
scenario, the TCB breaks up into SNBs. 
In the second scenario, no bifurcation is left in the presence of the 
perturbation and the functions tr\,M$(\epsilon)$ approach the critical 
line tr\,M = 2 without reaching it, so that one may speak of an {\bf 
avoided bifurcation}. These scenarios can be described by the extended 
normal forms given in \Sec{secnormunftcb}. We then also investigate
perturbations that fulfill the criteria \eq{libcond}, allowing for the
existence of TCBs in systems with or without any discrete symmetries.

\newpage

\subsubsection{Addition of a homogeneous transverse magnetic field}
\label{secbfield}

We first discuss the addition of a homogeneous magnetic field 
{\bf B = e$_z$}$B_0$ to the Hamiltonian \eq{ghh} which is transverse
to the $(x,y)$ plane of motion. This is a situation that is frequently
set up in experimental physics and gives us one important way of
breaking of the time-reversal symmetry. The momenta $p_i$ ($i=x,y$) in 
\eq{ghh} are replaced by the standard ``minimal coupling'',
\be
p_i \; \rightarrow \; p_i - \frac{e}{c}\,A_i\,, \qquad
{\bf A} = \frac12\,({\bf r}\times{\bf B})\,,
\ee
where ${\bf A}$ is the vector potential and $e$ the charge of the
particle. This adds the following perturbation to the Hamiltonian:
\be
\delta H(x,y,p_x,p_y) = \frac{eB_0}{2c}\,(xp_y-yp_x) 
                      + \frac12\left(\frac{eB_0}{2c}\right)^{\!2}(x^2+y^2)\,,
\label{bfieldpert}
\ee
which breaks the time-reversal symmetry of \eq{ghh} due to the 
linear terms in $p_x$ and $p_y$, but also breaks the straight-line
libration condition \eq{libcond}.

As an example, we choose the GHH potential with $\gamma=0.5$, $\beta=0.1$. 
Here the saddle energy for the A' orbit is $e_1=3.83$; the other
saddles are at $e_0=0.9852$ and $e_2=6.35$. In \fig{crossfig} we show 
the stability traces tr\,M$(e)$ of the orbits A' and B' with and without
magnetic field. For $B_0=0$ (triangles and dashed-dotted lines), these 
orbits A' and B cross at e$_{bif}=1.42665$ in a TCB like in the
examples discussed above. For $B_0\neq 0$ (circles and solid lines),
they rearrange themselves into pairs A'$_{\!4}$/B$_5$ and B$_4$/A'${\!_5}$ 
colliding in SNBs according to the prediction
\eq{trm2tb} of the normal form \eq{nf2tb}, in which $\kappa$ is taken
proportional to the value of $B_0$.

\Figurebb{crossfig}{40}{50}{767}{510}{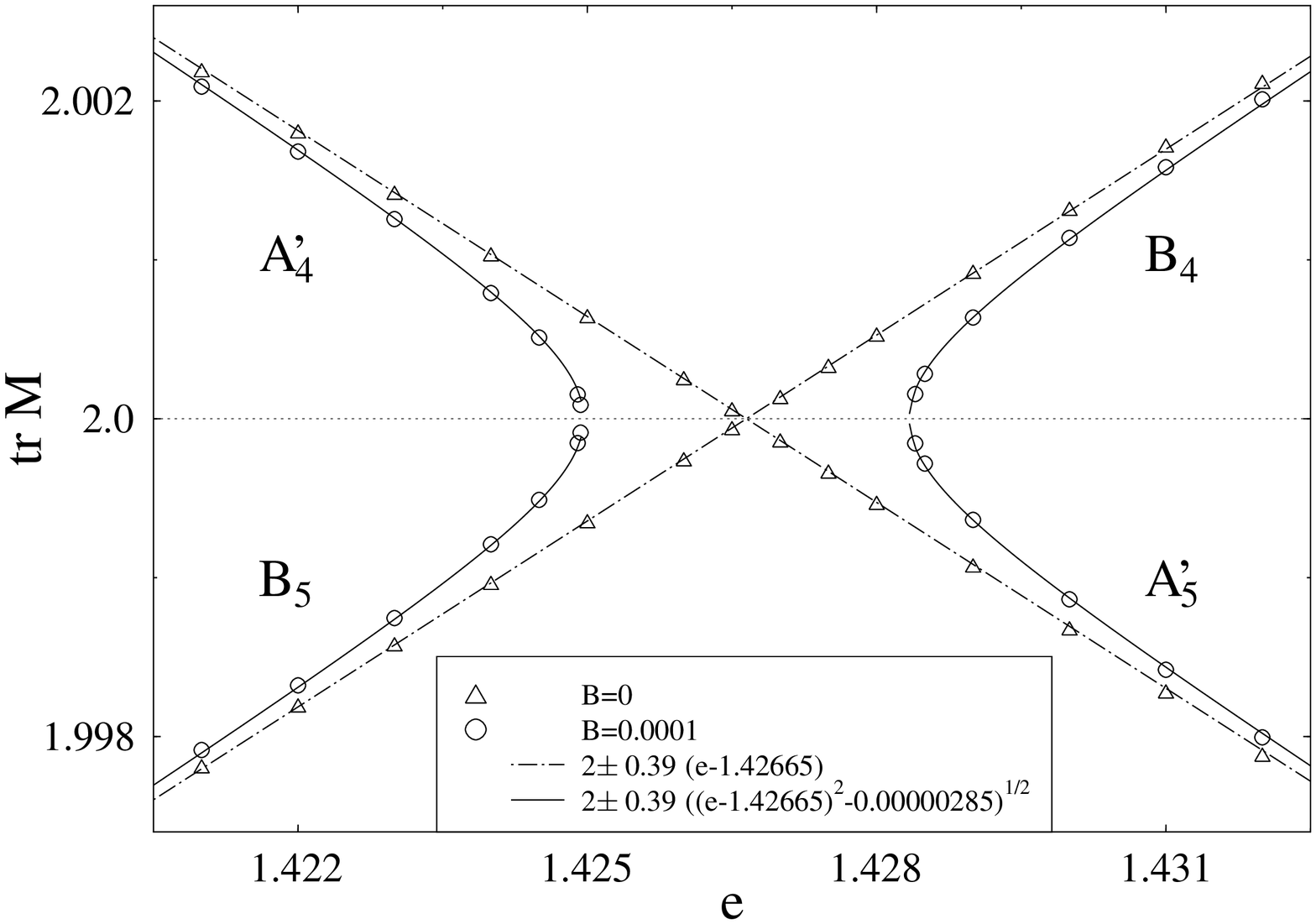}{8}{16.5}{
Unfolding of a TCB by a transverse magnetic field in the GHH potential 
with $\gamma=0.5$, $\beta=0.1$. Shown is tr\,M versus scaled energy $e$. 
{\it Dashed lines and triangles:} prediction \eq{trMn} and numerical 
results for field strength $B_0=0$ of the unperturbed TCB; 
{\it solid lines and circles:} local prediction \eq{trm2tb} (with 
an adjusted value of $\kappa$) and numerical results for $B_0=0.0001$.
}

\newpage

\subsubsection{Destruction of the TCB by a perturbation of the potential}
\label{secdespot}

Another example of the same unfolding of a destroyed TCB is shown in 
\fig{etafig}. Here the unperturbed GHH potential is the same as that
used in \fig{tcbdeg} above, which is symmetric about the $y$ axis.
This time we apply a perturbation of the potential alone
\be
\delta V(x,y) = \kappa\,x'y'^3\,,
\label{pertpot}
\ee
whereby $x',y'$ are rotated Cartesian coordinates such that the 
bifurcating A' orbit lies on the $y'$ axis.
Clearly, this perturbation does not fulfill the conditions
\eq{libcond} (expressed in the rotated coordinates) and in fact
destroys the original TCB of the orbits A and B' shown in \fig{tcbdeg};
the same fate happens also to the pair A'' and B'' of orbits. We see in
\fig{etafig} that, again, the original pairs of orbits on either
side of the unperturbed TCB rearrange themselves such as to destroy
each other in two pairs of SNBs, each according 
to the prediction \eq{trm2tb} of the corresponding normal form.
Since the effective perturbation strengths are different in 
the two original directions of the A' and A'' orbits, the splitting
between the two pairs of SNBs is slightly different.
A problem arises with the nomenclature of the perturbed orbits,
which is somewhat {\it ad hoc}, since all perturbed orbits have become
rotations. In the square brackets in the figure we indicate the names 
of the unperturbed orbits, of which A', A'' are straight-line and
B', B'' curved librations (their stability traces are shown in 
\fig{tcbdeg} above). The stability traces of the perturbed orbits
change drastically at the original bifurcations, but approach those
of the unperturbed orbits sufficiently far from the bifurcations.
The insert in the upper left of \fig{etafig} illustrates one possible
unfolding of a destroyed isochronous PFB (that seen at $e=0.34$ 
between the orbits B and B'/B'' in \fig{tcbdeg}) and will be 
commented in \Sec{secunfpfb} below.

\Figurebb{etafig}{20}{35}{783}{370}{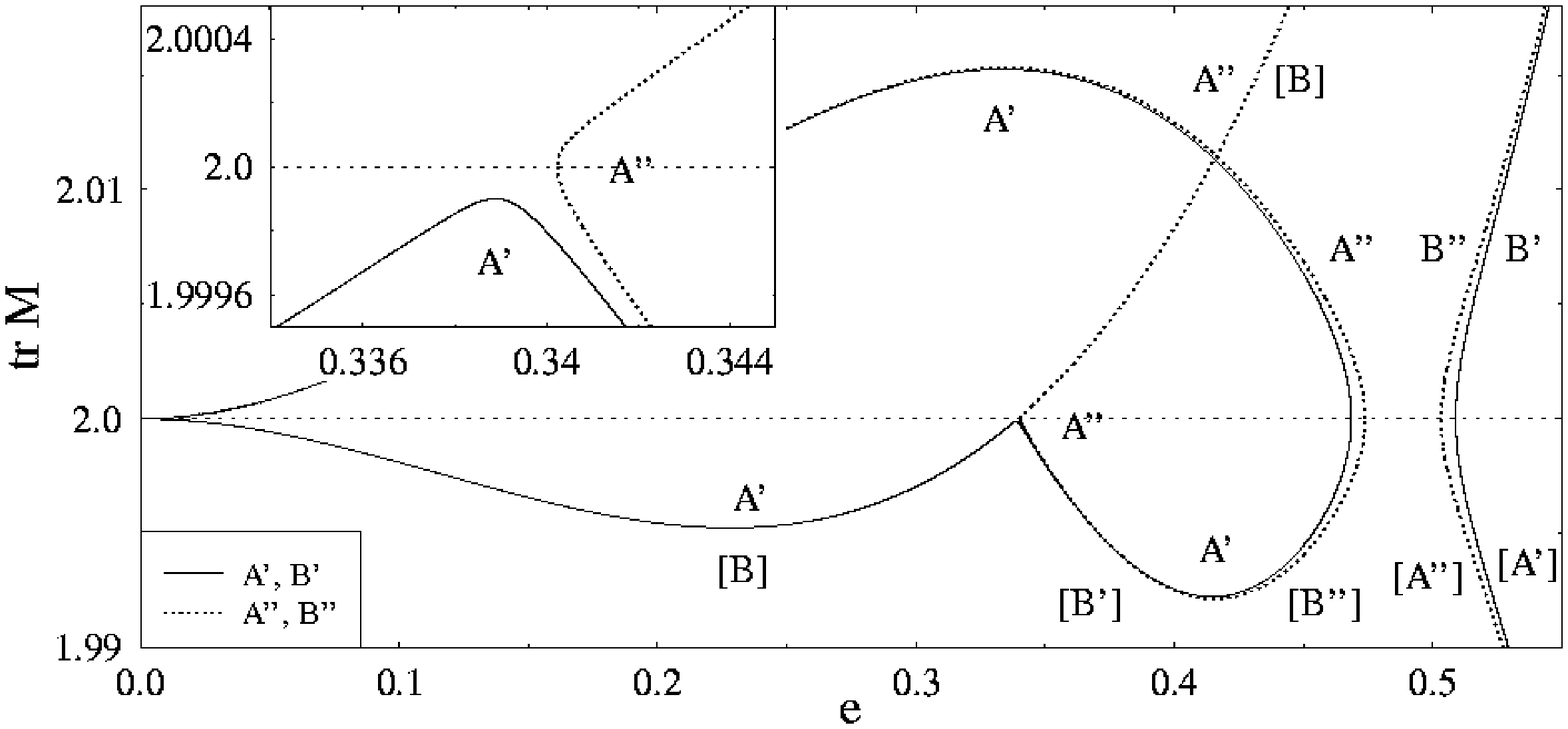}{6.5}{16.5}{
Unfolding of the TCB shown in \fig{tcbdeg} under the perturbation
\eq{pertpot} with $\kappa=0.0001$ (see text for details). The labels 
in brackets [\,] correspond to the orbits of the unperturbed system in
\fig{tcbdeg}. For the insert, see \Sec{secunfpfb}. 
}
\vspace*{-1.0cm}

\newpage

\subsubsection{An avoided TCB}
\label{secavoid}

In \fig{gamfig} we give an example of an avoided bifurcation. We
start again from the same example as in \fig{tcbdeg}, but now we
apply the following perturbation:
\be
\delta H(x,y,p_x,p_y) = \kappa'\,x'^2p_{y'}\,,
\label{pertcons}
\ee
again in the same rotated coordinates as for the perturbation \eq{pertpot} 
above. By construction, this perturbation does fulfill the
libration-conservation conditions \eq{libcond}, expressed in the
rotated coordinates ($x',y'$) for the orbit A', so 
that the TCB of the orbits A' and B' survives. It will be discussed
in more detail in the next subsection. The perturbation \eq{pertcons}
destroys, however, the TCB of the original orbit pair A'' and B''
at $e_{tcb}=0.489$ and is seen to lead to an avoided bifurcation of 
the perturbed orbits, which are now called A'' and B and shown by the 
heavy dashed lines. Their stability traces follow the local behavior 
\eq{trmavoid} predicted by the normal form \eq{nfavoid}. Again, our 
nomenclature for the new orbits is not strict; the perturbed B'' 
orbit has, for $e>0.489$, become a portion of the new orbit B. As in 
\fig{etafig}, the graphs tr\,M$(e)$ of the perturbed new orbits 
approach the unperturbed ones far from the bifurcations.

\vspace*{-0.5cm}
\Figurebb{gamfig}{20}{40}{719}{290}{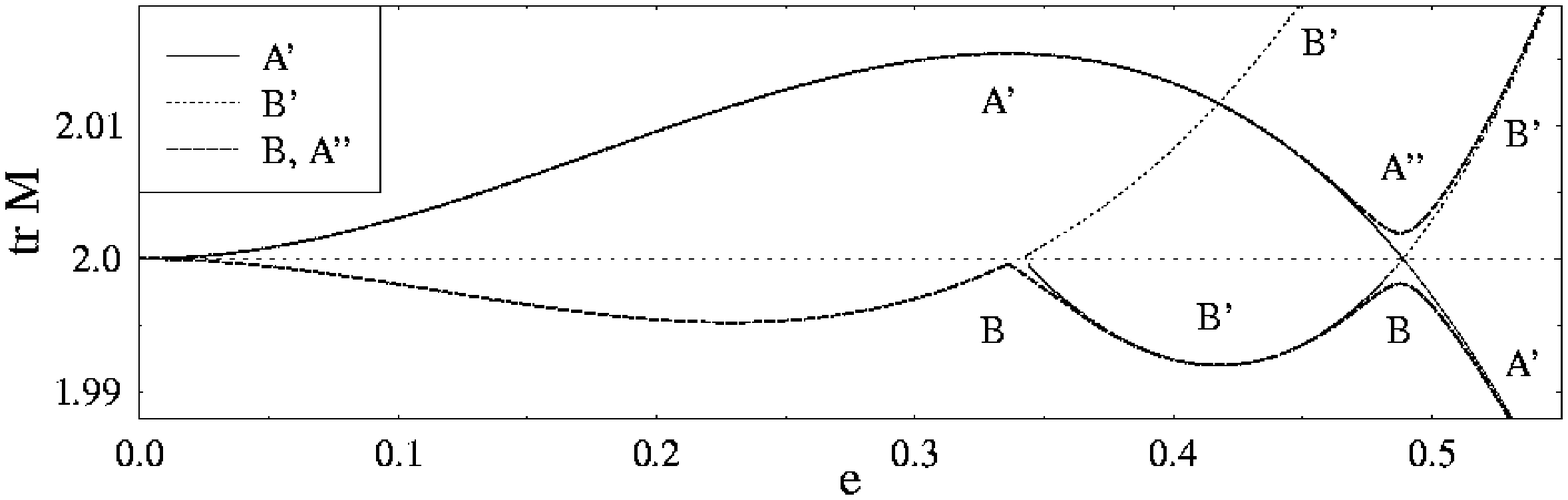}{4.5}{16.5}{
Unfolding of the TCB of the orbit pair A'' and B'' shown in
\fig{tcbdeg} under the perturbation \eq{pertcons} with 
$\kappa'=0.01$ into an avoided bifurcation of the new orbit
pair A'' and B near $e=0.489$ (shown by the heavy dashed lines).
The surviving TCB of the orbit pair A' and B' (solid and thin
dashed line, respectively) is commented in \Sec{secsurv}.
}

\subsubsection{TCB in a system without any discrete symmetry}
\label{secsurv}

We now come to our last, and perhaps most interesting, example:
a TCB in a system without any discrete symmetry. It is shown
in \fig{gamfig} by the solid line for the orbit A' and the
thin dashed line for the orbit B'. It is the same as the TCB
shown in \fig{tcbdeg} after applying the perturbation \eq{pertcons}
that has been explicitly constructed so as to preserve the
straight-line libration condition \eq{libcond} in the rotated
coordinates $x'$, $y'$. Here $y'$ is the
direction of the A' orbit. Thus, the libration A' in the perturbed
system is identical to that in the unperturbed GHH potential
($\gamma=0.75$, $\beta=0$). The orbit B', however, which in the
unperturbed GHH system is a curved libration similar to that
shown in \fig{cross0}, has now become a rotation except at the
TCB point. While it was originally created,
together with its symmetry-degenerate partner B'', in an 
isochronous PFB at $e=0.34$ from the original B orbit (see
\fig{tcbdeg}), this PFB is destroyed under the perturbation
\eq{pertcons}, and the perturbed B' orbit is now created at a 
SNB at $e=0.343$. Its stable lower branch is 
that which crosses the unchanged A' orbit transcritically at the 
slightly shifted new bifurcation energy $e_{tcb}=0.4886$. 

The shapes of this perturbed B' orbit in the rotated $(x',y')$
plane are shown in \fig{shapbxy}, on the left side in the energy 
region between its creation at $e=0.343017$ and its TCB at $e=0.4886$
where it is stable, and on the right side for the energies
$e\geq0.4886$ where it is unstable. Its librational shape at 
$e=0.4886$, where it is identical to the A' orbit, is shown in 
both panels of the figure (note their different scales!).

\Figurebb{shapbxy}{25}{25}{750}{490}{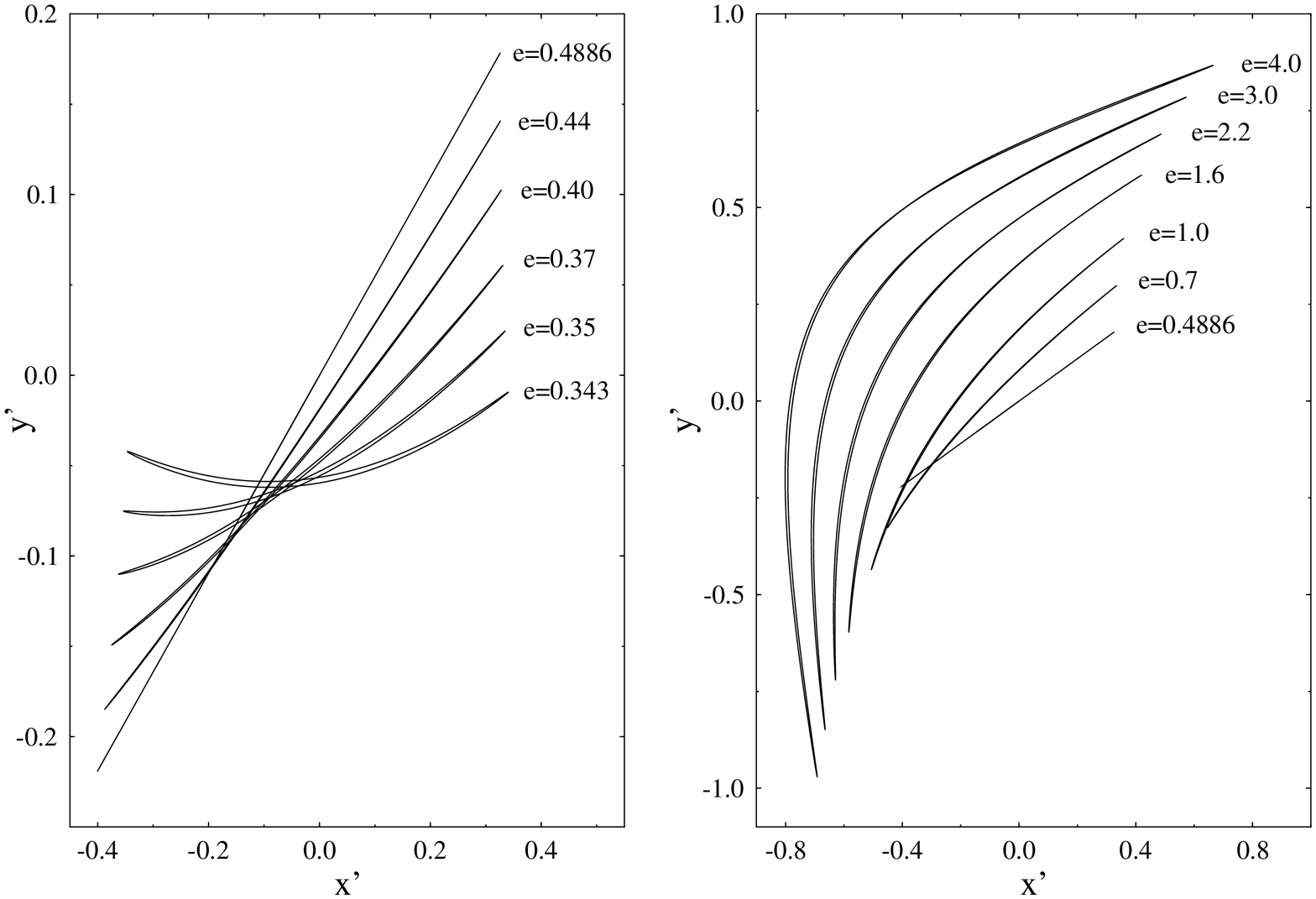}{10.8}{16.5}{
Shapes of the B' orbit, shown by the dashed line in \fig{gamfig},
in the rotated $(x',y')$ plane at different energies. {\it Left
panel:} stable region, at the energies $e=0.343$ (creation in 
SNB) -- 0.4886 (TCB point). {\it Right panel:} 
unstable region, at the energies $e=0.4886$ (TCB point) -- 4.0. 
Note the different scales in the two panels.
}

\vspace*{-0.7cm}
\Figurebb{shapbpxpy}{20}{10}{790}{350}{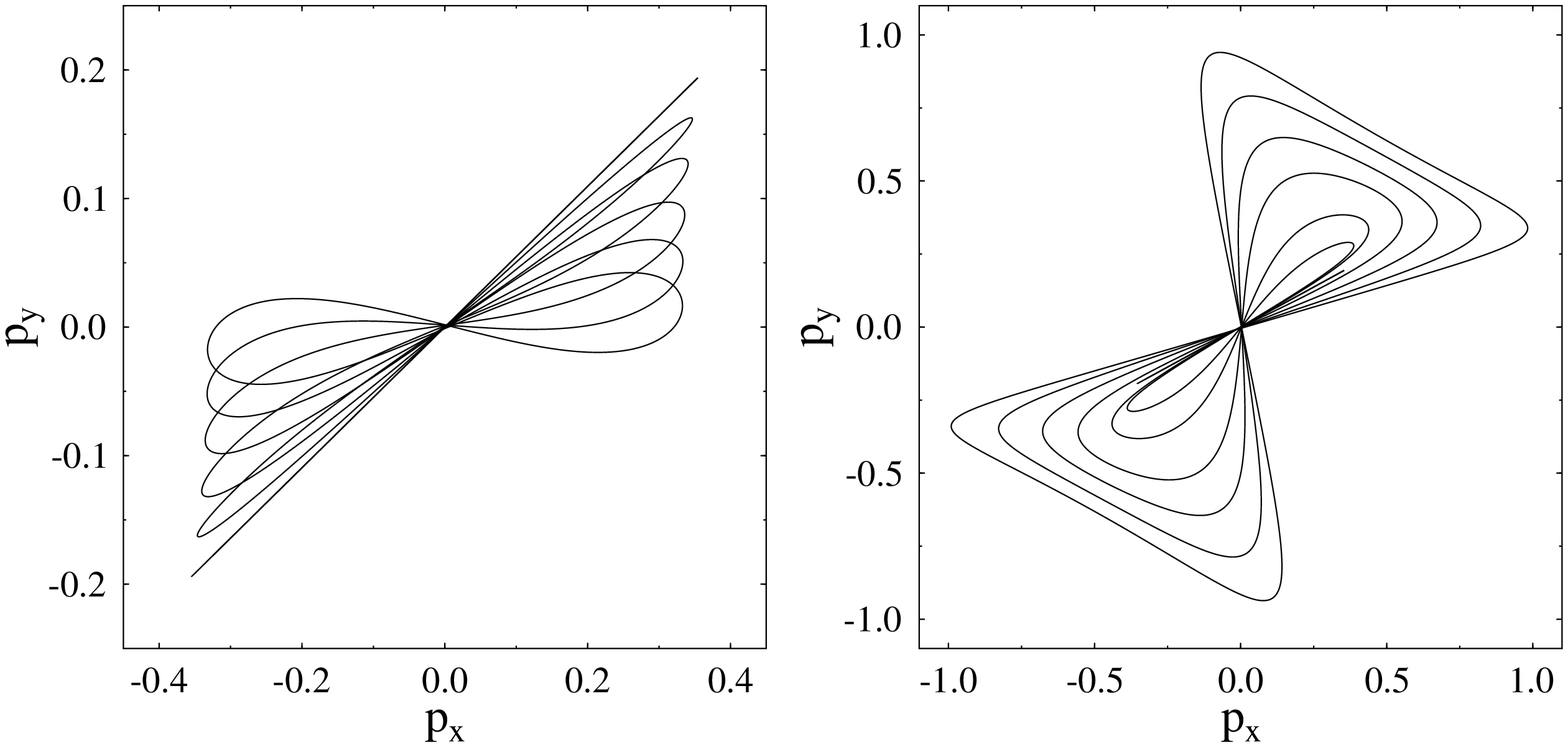}{6}{16.5}{
Same as \fig{shapbxy} but in the rotated momentum space 
$(p_{x'},p_{y'})$.
}
\vspace*{-0.5cm}
In \fig{shapbpxpy} we present the shapes of the B' orbit in the
rotated momentum space $(p_{x'},p_{y'})$. Here the orbit appears as
a figure-8 type rotation, except at the TCB where it must be a
straight-line libration, as the A' orbit, also in momentum space.
It should be noted that qualitatively, the shapes in momentum
space are the same for all transcritically bifurcating B type 
orbits discussed in this paper, even if they remain curved
librations in coordinate space.

One may interpret the perturbation \eq{pertcons} as the first-order
expansion of a weak {\bf inhomogeneous} magnetic field with
strength proportional to $\kappa'$. In a homogeneous field, for
which the full perturbation is given in \eq{bfieldpert}, the
Lorentz force tends to turn all straight-line orbits into curved 
librations. In the system perturbed by \eq{pertcons}, the Lorentz
force of this inhomogeneous magnetic field is canceled, at least to 
lowest order, by the geometry of the total potential which tends to 
curve the librations the other way round. We leave it to the 
interested reader to speculate whether this scenario finds 
applications in accelerator physics, where one may want to produce 
straight-line trajectories in an inhomogeneous magnetic field.


\subsubsection{Creation of a TCB in the unfolding of a PFB}
\label{secunfpfb}

In this section we will show how a TCB can be created by perturbing a 
PFB and how its existence may depend on particular symmetries. Two 
characteristic unfoldings of isochronous PFBs in one-dimensional 
dynamical systems have been described in \cite{bad1,bad2}, which
correspond to the ``universal unfolding'' of the isochronous PFB 
discussed extensively in \cite{gols}. We find the same unfoldings for 
the isochronous PFBs of the straight-line librations in the (G)HH 
potentials; one of them is of particular interest here as it leads to 
a TCB. The corresponding normal forms are given in \App{secnormunfpfb}.

In the first scenario, the original parent orbit does not change its 
stability, thus avoiding the bifurcation, and a pair of new orbits 
is created at a SNB. One of these new orbits takes the role of the 
original parent orbit after the bifurcation, and the perturbed parent 
orbit takes the role of one of the new orbits created at the original 
PFB. Examples of this scenario can be seen in \fig{etafig} (inserted 
close-up) and in \fig{gamfig}, as results of two different 
perturbations of the same original PFB seen in \fig{tcbdeg} at $e\sim 0.34$.

The second scenario is the unfolding into a SNB of a pair of new orbits, 
followed by a TCB of one of these orbits with the original parent orbit. 
An example of this has already been pointed out in \fig{trace} (left 
panel) to occur near $e\sim 1.62$, where the orbits B' and B'' 
bifurcate from the A'' orbit. In the following we shall further 
illustrate this unfolding by explicitly perturbing a PFB in the 
standard HH system.

We start from the HH system, i.e., \eq{ghh} with $\gamma=1$, $\beta=0$, 
and add the following perturbation to the potential
\be
\delta V(x,y) = \frac13\,\delta\,x^3,
\label{pertpfb}
\ee
which destroys both the $C_{3v}$ symmetry and the reflection symmetry 
at the $y$ axis. It therefore affects the cascade of isochronous 
PFBs of the linear A orbit along the $y$ axis (cf. Sec \ref{secshh}). 
The perturbation \eq{pertpfb} is chosen such as to preserve the
straight-line libration condition \eq{libcond}, so that the A orbit
still exists in its presence. To ensure the presence of a TCB in the
perturbed system, we must fulfill the condition $P_{qq}\neq0$ given in 
\eq{extcl}. An explicit expression for the quantity 
$P_{qq}$ in terms of the (total, perturbed) potential $V$ is given in 
\eq{Pqq}. [In the integrand of \eq{Pqq}, the function $V_{xxx}(x,y)$ 
is taken along the A orbit with $x(t)=0$, $y=y_A(t)$; see \Sec{seclib} 
for details and notation.] Since $V_{xxx}$ becomes nonzero 
with the perturbation \eq{pertpfb}, the occurrence of a TCB is possible.
But $V_{xxx}\neq 0$ is not sufficient to ensure $P_{qq}\neq 0$: this
will also depend on the symmetry of the function $\xi_1(t)$ appearing
in the integrand of the quantity $P_{qq}$ in \eq{Pqq}.   

Now, as discussed in \cite{lame,mbgu}, the functions $\xi_1(t)$
describe the $x$ motion (transverse to the A orbit) of the new
orbits created at the successive PFBs of the A orbit. These functions
are periodic Lam\'e functions with well-known symmetry properties. 
As it turns out, $\xi_1(t)$ of the L type orbits born
at every second PFB of the cascade are even functions of $t$, where
$t=0$ is the time at which $y_A(t)$ is maximum; whereas those of the
R type orbits born at every other bifurcation are odd. The result is
that $P_{qq}$ becomes zero at the R type bifurcations, in spite of
$V_{xxx}\neq 0$, while  $P_{qq}\neq0$ for the L type bifurcations.
Consequently, it is only at the L type bifurcation energies that a
TCB can exist in the perturbed system. 
Our numerical investigations have confirmed that under the perturbation
\eq{pertpfb} all R type bifurcations remain, indeed, unbroken PFBs 
with unchanged stability traces tr\,M$(e)$ to first order in $\delta$,
while the L type bifurcations are broken up as discussed above.

In \fig{bifl6} we show the creation of the orbits L$_6$ and L'$_6$
from the A orbit in the HH system under the perturbation \eq{pertpfb}.
In the unperturbed HH system, these orbits are born as a degenerate 
pair from a PFB at the energy $e_6=0.986709235$ (cf. \cite{lame}), as 
also seen in \fig{hhtr}. Here the PFB has been broken according to the
second scenario described above, unfolding into a TCB of A$_{6/7}$ and 
L$_{7/6}$ at precisely the same critical energy $e_{tcb}=e_6$, and a 
SNB at $e_6-\Delta e_6\sim 0.98703$ where L$_7$ and L'$_6$ are born. 
The thin dash-dotted line gives the slope of tr\,M$(e)$ of the L orbit 
at the TCB which is minus that of the A orbit, as is characteristic of 
a TCB. The thin dotted line gives the slope of the original degenerate 
pair L$_6$, L'$_6$ created in the unperturbed HH system; this slope is 
minus twice that of the parent A orbit, as is typical of a PFB [see 
\eq{slopesfb} in \Sec{secpfb}]. The same scenario is found at all 
successive L type bifurcations. In the perturbed system, the orbit 
pairs L, L' are no longer degenerate, since the reflection symmetry at 
the $y$ axis is broken by \eq{pertpfb}. The graphs tr\,M$(e)$ seen in 
this figure are predicted by the normal form \eq{unfpfb0}, as
discussed explicitly in \App{secnormunfpfb}.

\Figurebb{bifl6}{20}{10}{730}{450}{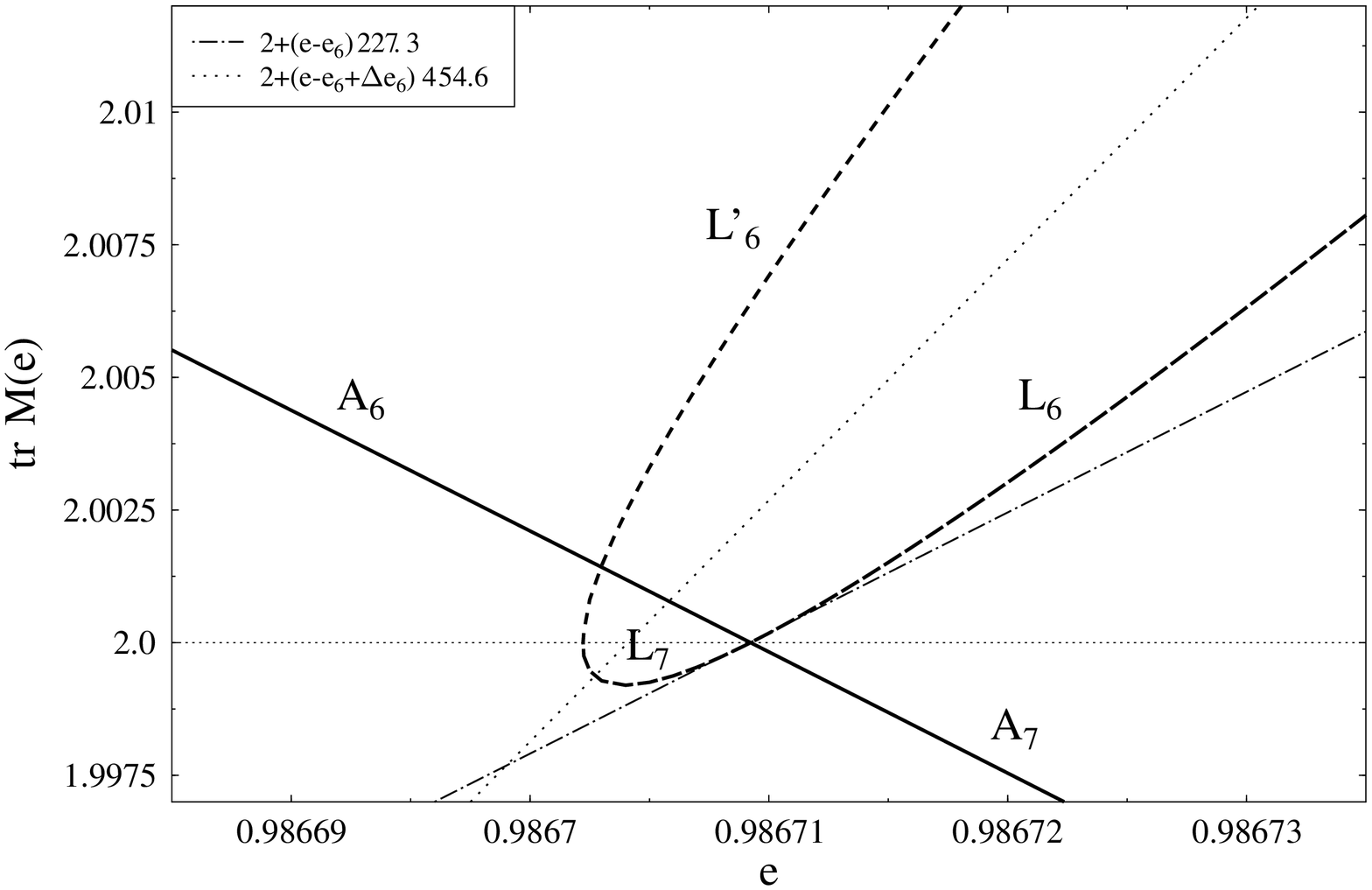}{8.5}{16.5}{
Creation of orbits L$_6$, L'$_6$ in a broken PFB in the HH 
potential under the perturbation \eq{pertpfb} with $\delta=0.5$. The
TCB of the orbits A and L occurs at the energy $e_6=0.986709235$ of
the PFB of the degenerate pair L, L' in the unperturbed HH 
potential \cite{lame}. (See text for the thin lines.)
}

\newpage

\section{Inclusion of a TCB in the semiclassical trace formula}
\label{secscl}

\subsection{The semiclassical trace formula}
\label{sectf}

Our investigations have been largely motivated by the use of
periodic orbits in the semiclassical description of the density 
of states $g(E)$ in a quantum system with discrete spectrum 
$\{E_i\}$ 
\be
g(E) = \sum_i \delta(E-E_i)\,.
\label{dos}
\ee
Initiated by Gutzwiller (see \cite{gutz} and earlier references
therein), the periodic orbit theory (POT) \cite{chaos} (see also 
\cite{stoe,book,gutzh} for general introductions) states that the
oscillating part of the quantum density of states is given, to 
leading order in $\hbar$, by a semiclassical trace formula of the 
form 
\be
\delta g(E) = \sum_{po}{\cal A}_{po}(E)\,\cos\left(\frac{S_{po}(E)}{\hbar}
                                       -\frac{\pi}{2}\,\sigma_{po}\right).
\label{tf}
\ee
The sum goes over all periodic orbits ($po$) of the classical
system (including their repetitions), $S_{po}(E)$ are their actions 
\eq{action} and $\sigma_{po}$ their Maslov indices (see 
\cite{gutz,crl,sugi,mpmb} and \cite{book}, appendix D, for details). 
The ${\cal A}_{po}(E)$ are semiclassical amplitudes which depend on 
the nature of the orbits. For orbits that are {\bf isolated} in phase
space, the amplitudes were given by Gutzwiller \cite{gutz} as
\be
{\cal A}_{po}(E) = \frac{1}{\pi\hbar}\,\frac{T_{ppo}(E)}
                   {\sqrt{|{\rm det\,[M}_{po}(E)-{\rm I}]|}}\,,
\label{ampiso}
\ee
in terms of their primitive periods $T_{ppo}(E)=$ d$S_{ppo}(E)$/d$E$ 
and stability matrices M$_{po}(E)$; here I is the unit matrix with the
same dimension as M$_{po}$. For systems with continuous symmetries, 
in which most periodic orbits come in degenerate families (in particular
in integrable systems), explicit expressions for the ${\cal A}_{po}(E)$ 
have been derived by various authors \cite{cont}.
 
One problem with the Gutzwiller trace formula in mixed systems, where 
stable and unstable periodic orbits coexist, is the divergence of the 
amplitudes \eq{ampiso} occurring at bifurcations. Remedy is given by 
uniform approximations, introduced by Ozorio de Almeida and Hannay 
\cite{ozha} (see also \cite{ozob}) and further developed by several 
authors both for codimension-one \cite{ssu1,sun,ssu2} and codimension-two 
bifurcation scenarios \cite{kaid,scho}. In the following sections, we 
will discuss the uniform approximations and derive its appropriate 
form for a pair of transcritically bifurcating orbits.

The trace formula \eq{tf} does not converge in mixed systems and most
chaotic systems, in which the number of periodic orbits proliferates
exponentially with increasing length, so that the summation over all  
orbits typically cannot be performed (see \cite{gubu,chaos}). In our 
study, we coarse-grain the density of states by convolution with a 
normalized Gaussian with width $\Delta E$, so that only the shortest 
orbits with periods $T_{po}\siml\hbar/\Delta E$ contribute to the sum 
\cite{book,sist,sich}. Although the finer details of the spectral
information hereby are averaged out, the {\bf coarse-grained density 
of states} 
\be
g_{\Delta E}(E) = \frac{1}{\sqrt{\pi}\Delta E}
                  \sum_i e^{-(E-E_i)^2\!/\Delta E^2}
\label{doscg}
\ee 
still exhibits its gross-shell structure, provided that $\Delta E$ 
is not chosen too large. The correspondingly coarse-grained trace
formula reads \cite{book}
\be
\delta g_{\Delta E}(E) = \sum_{po}{\cal A}_{po}(E)\,
                         e^{-[T_{po}\,\Delta E/2\hbar]^2} 
                         \cos\left(\frac{S_{po}(E)}{\hbar}
                         -\frac{\pi}{2}\,\sigma_{po}\right)\!,
\label{tfcg}
\ee
where it can be seen that the additional exponential factor suppresses 
the contribution of longer orbits. This version of the POT has found 
many applications to gross-shell effects in finite fermion systems 
(see \cite{book,strum} for examples).

\subsection{Uniform approximation for bifurcating orbits}
\label{secunifo}

In this section we sketch the derivation for the combined contribution
of a pair of bifurcating orbits A and B to the semiclassical 
trace formula \eq{tf} for the density of states. Since the individual
amplitudes in the form \eq{ampiso} given by Gutzwiller diverge at the
bifurcation, one has to go one step back in their evaluation and
transform the trace integral to the phase space \cite{ssu1,crli1,magn}.
After doing the integration along the primitive A orbit\footnote{recall 
that we only consider primitive ``period one'' orbits and their isochronous 
bifurcations here.} with action $S_A(E)$, the remaining part of the trace 
integral is over the Poincar\'e surface of section in the variables
$Q$ and $p$ transverse to the A orbit:
\be
\delta g(E)\; = \; \Re e\;
                        e^{\frac{i}{\hbar}S_A(E)-i\frac{\pi}{2}\sigma_A}
                        \!\int \!{\rm d} Q \int \!{\rm d} p \; C(Q,p,\epsilon)\,
                        e^{\frac{i}{\hbar}{\widetilde S}(Q,p,\epsilon)}.
\label{phsptf}
\ee
The action function in the phase of the integrand is given by
\be
{\widetilde S}(Q,p,\epsilon) = {\widehat S}(Q,p,\epsilon) 
                             - S_A(\epsilon) - Qp\,,
\label{sphase}
\ee
where ${\widehat S}(Q,p,\epsilon)$ is the {\bf generating function} of
the canonical transformation \eq{QPofqp} that describes the Poincar\'e
map, and $S_A(\epsilon)$ is the action integral \eq{action} of the A 
orbit as a function of the control parameter $\epsilon$. By virtue of 
the canonical relations of the generating function
\be
P = \papa{{\widehat S}}{Q}\,, \qquad q = \papa{{\widehat S}}{p}\,,
\label{canonrel}
\ee
the stationary condition of the function ${\widetilde S}$ in the
$(Q,p)$ plane for any fixed $\epsilon$,
\be
\papa{{\widetilde S}}{Q}(Q_0,p_0,\epsilon) 
     = \papa{{\widetilde S}}{p}(Q_0,p_0,\epsilon) = 0\,,
\ee
yields $P_0=p_0$ and $Q_0=q_0$, so that the stationary points
$(Q_0,P_0,\epsilon)=(q_0,p_0,\epsilon)$ of the phase function 
\eq{sphase} are the fixed-point branches of the map and hence 
correspond to the periodic orbits. [Note that, by construction, 
${\widetilde S}(q_0,p_0,\epsilon)=0$ along the fixed-point branch 
of the A orbit.]

The amplitude function $C(Q,p,\epsilon)$ in \eq{phsptf} is given 
\cite{ssu1} in terms of the generating function 
${\widehat S}(Q,p,\epsilon)$ by
\be
C(Q,p,\epsilon) = \frac{1}{2\pi^2\hbar^2}\,
                  \papa{{\widehat S}}{E}(Q,p,\epsilon)\,.
\label{camp}
\ee
Note that
\be
\papa{{\widehat S}}{E}(Q,p,\epsilon) =: {\widehat T}(Q,p,\epsilon)
     = T_A(E) + \papa{{\widetilde S}}{E}(Q,p,\epsilon)\,,
\label{that}
\ee
where $T_A(E)=\frac{\rm d}{{\rm d}E}S_A(E)$ is the period of the 
A orbit. 

In principle, the integration over $Q$ and $p$ in \eq{phsptf} is 
limited to 
that domain of the $(Q,p)$ plane which is accessible under energy
conservation. However, in the spirit of the stationary-phase
approximation (including its extensions below) we expect that,
due to the rapidly oscillating phase of \eq{phsptf} in the
semiclassical limit ${\widetilde S}\gg\hbar$, the main
contributions to the integral come from small regions around the 
stationary points of the function ${\widetilde S}(Q,p,\epsilon)$. 
Assuming that the fixed points of A and the other 
orbit(s) taking part in the bifurcation are situated in the 
interior of this domain, and that no other bifurcations happen
nearby, we may extend the integrals over both $Q$ and $p$ from
$-\infty$ to $+\infty$.

Sufficiently far away from the bifurcation point $\epsilon=0$,
so that the orbits A and B are isolated, the stationary-phase
integration of \eq{phsptf} will yield precisely the contributions 
of the isolated orbits A and B to the standard Gutzwiller trace 
formula \eq{tf} with their individual amplitudes \eq{ampiso}.
Near the bifurcation, the stationary-phase approximation fails and
one has to include higher than second-order terms in $Q$ and $p$ in
the function ${\widetilde S}(Q,p,\epsilon)$. The simplest solution
\cite{ozob} is to use a truncated Taylor expansion of 
${\widetilde S}(Q,p,\epsilon)$ in all three variables, keeping only 
the minimum number of terms necessary to be able to reproduce locally 
the fixed-point branches of all the orbits taking part in a given 
bifurcation. These truncated forms of ${\widetilde S}(Q,p,\epsilon)$ 
are the {\bf normal forms} which are discussed in \App{nofosec}.

\subsection{Uniform approximation for the TCB}

Equipped with the normal form of the TCB given in \App{secnormtcb}, 
we now calculate the contribution of a pair of periodic orbits A and B 
undergoing a TCB to the semiclassical trace formula. We follow closely 
the treatment of \cite{ssu1}, where uniform approximations for the
generic bifurcations corresponding to \cite{meye,ozob} were derived.

Since \eq{phsptf} is 
invariant under canonical transformation $(Q,p)\rightarrow (Q',p')$, 
we may think of the variables $Q,p$ to be the adapted coordinates for 
which \eq{split} and the equations given 
thereafter are valid. We are therefore allowed to insert for 
${\widetilde S}(Q,p,\epsilon)$ the normal form derived 
in \Sec{secnormtcb} for the TCB, in order to derive the 
uniform approximation to the trace formula which includes the orbits 
taking part in the TCB. 

We will do this in two steps. First, we evaluate \eq{phsptf} only
at $\epsilon=0$. This yields the so-called {\bf local uniform 
approximation} in the spirit of Ozorio de Almeida and Hannay 
\cite{ozha}. In the second step, we use the full normal form 
\eq{nftcb} and the corresponding functions $C(Q,p,\epsilon)$ defined 
by \eq{camp} to find, after some suitable transformations, the global 
uniform approximation in the spirit of \cite{ssu1,sun,ssu2}. The 
latter yields asymptotically the Gutzwiller trace formula for the 
orbits A and B sufficiently far from the bifurcation.

We now use for ${\widetilde S}(Q,\epsilon)$ the normal form 
\eq{nftcbtil} of the TCB (omitting the tilde on the variables 
$Q,p,\epsilon$ and on $b$). Using the relation \eq{eps}, the function 
${\widehat T}$ in \eq{that} becomes
\be
{\widehat T}(Q,\epsilon) = T_A(E) - Q^2\,.
\ee
After the elementary $p$ integration, yielding a complete Fresnel
integral, we obtain for \eq{phsptf}
\be
\delta g(E) \; = \; \frac{1}{\pi\hbar\sqrt{2\pi\hbar}}\, \Re e\;
                         e^{\frac{i}{\hbar}S_A(E)-i\frac{\pi}{2}(\sigma_A+\frac12)}
                         \left[T_A(E)\,F(b,\epsilon)-G(b,\epsilon)\right],
\label{tfFG}
\ee
where we have defined the two following one-dimensional integrals
\bea
F(b,\epsilon) & := & \int_{-\infty}^\infty\!{\rm d} Q\,
                     e^{-\frac{i}{\hbar}\,(\epsilon Q^2+bQ^3)},\label{Fint}\\
G(b,\epsilon) & := & \int_{-\infty}^\infty\!{\rm d} Q\,Q^2\,
                     e^{-\frac{i}{\hbar}\,(\epsilon Q^2+bQ^3)}\label{Gint}
                 = i\hbar\,\papa{}{\epsilon}F(b,\epsilon)\,. 
\eea
Using the substitution $Q=x-{\tilde Q}$ with ${\tilde Q}=\epsilon/3b$,
we obtain
\be
F(b,\epsilon) = 2\pi\left(\frac{\hbar}{3|b|}\right)^{\!1/3}
                e^{\frac{i}{\hbar}\Delta S(\epsilon)}\,{\rm Ai}(-z')\,,
                \qquad
           z' = \frac{\epsilon^2}{(3|b|)^{4/3}\hbar^{2/3}}\,,
\label{Fb}
\ee
where Ai is the Airy function (see \cite{abro}, 10.4) and
\be
\Delta S(\epsilon) = -\frac{2\epsilon^3}{27\,b^2}\,.
\label{DelS}
\ee
Using the r.h.s.\ of \eq{Gint} to calculate $G(b,\epsilon)$, we finally
obtain for the level density
\bea
\delta g(E) & \! = \! & \frac{\sqrt{2}}{\sqrt{\pi}\,\hbar^{7/6}\,(3|b|)^{1/3}}\, 
                       \Re e\; e^{i\left[\frac{1}{\hbar}S_A(E)+\Delta S(\epsilon)
                                         -\frac{\pi}{2}(\sigma_A+\frac12)\right]}\nonumber\\
            & &        \hspace*{3.2cm}
                       \times\left[\left(T_A(E)+\frac{2\epsilon^2}{9b^2}\right){\rm Ai}(-z')
                       +i\,\frac{2\epsilon}{(3|b|)^{4/3}}\,\hbar^{1/3}{\rm Ai}'(-z')\right]\!.
\label{tfuni1}
\eea

\subsubsection{Local uniform approximation}

We first give the result \eq{tfuni1} for $\epsilon=0$, using the known
value \cite{abro} of Ai(0), to find the local uniform level density at 
the bifurcation energy $E_0$:
\be
\delta g_{loc}(E_0) = \frac{T_A(E_0)\,\Gamma(\frac13)}
                      {\pi\sqrt{6\pi}\,\hbar^{7/6}\,|b|^{1/3}}\,
                      \cos\left[\frac{1}{\hbar}S_A(E_0)
                      -\frac{\pi}{2}\,\sigma_A-\frac{\pi}{4}\right]\!,  
\label{dgloc}
\ee
which contains the combined contribution of both orbits A and B
taking part in the transcritical bifurcation. An explicit expression 
for the calculation of the normal form parameter $b$ is given in 
\eq{Pqq} of \Sec{seclib}. The result \eq{dgloc} 
looks identical to that obtained in \cite{ssu1} for the generic 
SNB. The reason is that the normal form for this bifurcation is 
\cite{ssu1} $\widetilde{S}(Q,p,\epsilon)=-\epsilon\,Q-
b\,Q^3-\sigma\,p^2/\!2$ which for $\epsilon=0$ gives, of course, the 
same result as the normal form \eq{nftcbtil}. Note that the power 
7/6 of $\hbar$ in the denominator is by 1/6 higher than in 
the semiclassical amplitude \eq{ampiso} of an isolated orbit.

\subsubsection{Global uniform approximation}

The result \eq{dgloc} gives the correct semiclassical amplitude of
the bifurcating pair of orbits A and B only locally at the
bifurcation, i.e., for $\epsilon=0$. We want, however, to know it
also away from the bifurcation, and in particular, also in the
limit where it can be written as a sum of the two individual
contributions of the isolated orbits A and B to the standard 
Gutzwiller trace formula \eq{tf}. To achieve this, we note that 
if we use the asymptotic forms of the Airy function and its 
derivative in \eq{tfuni1} for $|z'|\gg 1$, we obtain two terms
that formally look like contributions to \eq{tf} with amplitudes
of the form \eq{ampiso}, but with the actions $S_{po}(E)$, 
periods $T_{po}(E)$, and stability traces tr\,M$_{po}(E)$
replaced by their expansion to lowest order in $\epsilon$, as 
found from the normal form and given in \eq{trMn} and 
\eq{delSntcb}. 

The next intuitively obvious step is therefore to rewrite the
asymptotic form of \eq{tfuni1} in terms of the locally expanded
quantities $S_{po}(\epsilon)$, $T_{po}(\epsilon)$, and 
tr\,M$_{po}(\epsilon)$ of the two orbits ($po$ = A,B) and then to 
replace them by the correct functions $S_{po}(E)$, $T_{po}(E)$, and 
tr\,M$_{po}(E)$ found numerically for the isolated orbits away
from the bifurcation. This step
has been rigorously justified in \cite{ssu1} by some appropriate
transformations and need not be repeated here. The calculation
goes exactly like that presented in \cite{ssu1} for the case of 
the SNB on that side where both orbits are real.
The reason is that although the normal forms of the two
bifurcations are different, they lead to identical integrals
after a translation in the integration variable $Q$.

The result is the following uniform contribution of the two 
bifurcating orbits to the Gutzwiller trace formula:
\bea
\delta g_{un}(E) = \sqrt{6\pi\xi}\left\{\frac{\overline{\cal A}(E)}{\sqrt{z}}\,
                \cos\left(\frac{\overline S(E)}{\hbar}
                          -\frac{\pi}{2}\,{\overline\sigma}\right)\!
                {\rm Ai}(-z) - \frac{\Delta{\cal A}(E)}{z}\,
                \sin\left(\frac{\overline S(E)}{\hbar}
                    -\frac{\pi}{2}\,{\overline\sigma}\right)
                {\rm Ai}'(-z)\right\}.
\label{dguni}
\eea
The quantities occurring in \eq{dguni} are defined as
\bea
                  z & = & (3\xi/2)^{2/3}\,, \qquad \qquad \quad\,
             \xi \;\; = \;\; \frac{1}{2\hbar}\,|S_A-S_B|\,,\nonumber\\
      {\overline S} & = & \frac12\,(S_A + S_B)\,, \qquad \quad\;\;
{\overline\sigma}\;\; = \;\;\frac12\,(\sigma_A+\sigma_B)\,,\nonumber\\ 
{\overline{\cal A}} & = & \frac12\,({\cal A}_A+{\cal A}_B)\,,\qquad\;
  \Delta{\cal A} \;\; = \;\;\frac12\,({\cal A}_A-{\cal A}_B)\,
                        {\rm sign}(S_A-S_B)\,, 
\eea
all to be taken at the energy $E$, where $S_{po}(E)$ and $\sigma_{po}$
are the actions and Maslov indices, respectively, of the isolated
periodic orbits on either side of the bifurcation, and ${\cal
 A}_{po}(E)>0$ are their Gutzwiller amplitudes \eq{ampiso}.

At the bifurcation ($\epsilon=z=\xi=0$), the result \eq{dguni} reduces
to the local uniform approximation \eq{dgloc}. Far enough away
from the bifurcation, it goes over to the contribution of the isolated 
orbits A and B to the standard Gutzwiller trace formula. Indeed, 
expressing the Airy function in terms of Bessel functions as \cite{abro}
\bea
{\rm Ai}(-z)  & = & \frac13\,\sqrt{z}\left[J_{1/3}(\xi)+J_{-1/3}(\xi)\right],
                    \nonumber\\
{\rm Ai}'(-z) & = & \frac13\,z\left[J_{2/3}(\xi)-J_{-2/3}(\xi)\right],
\eea
and using their asymptotic form
\be
J_\nu \longrightarrow \sqrt{\frac{2}{\pi\xi}}\,\cos\left(\xi-\frac{\pi}{2}\,\nu
                                                            -\frac{\pi}{4}\right)
\qquad \hbox{ for } \qquad \xi\gg 1\,,
\ee
we obtain from \eq{dguni} for $\xi\gg1$, i.e., for $|S_A-S_B|\gg 2\hbar$,
the sum of the isolated Gutzwiller contributions to the trace formula
\be
\delta g(E) = \sum_{po=A,B}{\cal A}_{po}(E)\,\cos\left(\frac{S_{po}(E)}{\hbar}
                                       -\frac{\pi}{2}\,\sigma_{po}\right),
\label{dgisoab}
\ee
with the amplitudes ${\cal A}_{po}(E)$ given in \eq{ampiso}.

For the reason given above, the result \eq{dguni} looks identical to
that given in \cite{ssu1} for the SNB on that side
where the two orbits are real. The present result holds on both sides
of the TCB and can easily be seen to yield asymptotically the result 
\eq{dgisoab} on both sides, with the roles of the orbits A and B
and their Maslov indices properly exchanged.


\subsection{Numerical test}
\label{sectesttf}

We present here a numerical calculation of the density of states, 
both quantum-mechanical and semiclassical, of the GHH Hamiltonian 
\eq{ghh} with $\gamma=0.6$, $\beta=0.07$, whose shortest orbits and 
stability traces are shown in \fig{trace}. The parameter $\alpha$ in
the Hamiltonian \eq{ghh} has been chosen as $\alpha=0.04$. The three
saddles then lie at the energies $E_0\simeq 103$, $E_1\simeq 293$ and
$E_2\simeq 390$ in units such that the spacing of the harmonic-oscillator 
spectrum reached in the limit $E\to0$ equals $\hbar\omega=1$. (See 
\fig{trace} for the scaled energies $e_0,$ $e_1$ and $e_2$ of the 
saddles.) 

The spectrum of the quantum-mechanical Hamiltonian corresponding to 
\eq{ghh} has been obtained by diagonalization in a two-dimensional 
harmonic oscillator basis. Strictly speaking, the spectrum is not 
discrete since the system has no lower bound. However, for energies 
below the three saddles, the tunneling probabilities are exponentially 
small. In principle, the semiclassical trace formula can also be 
applied in the continuum region above the saddles, if the (complex)
energies of the resonances are used to calculate the density of states. 
For a detailed discussion of this situation, we refer to \cite{kwb} 
where a semiclassical calculation has been successfully performed for 
the standard HH system up to twice the saddle energy. In the present 
system, the discrete energies obtained for $E_0\siml E \siml 150$
in the numerical diagonalization turn out to be good approximations 
to the real parts of the resonance energies, while the imaginary parts 
of the resonances are still negligible.

For our present test, we have chosen a coarse-graining width 
in \eq{doscg} of $\Delta E = 0.6$. This allows us to restrict the
summation over the periodic orbits ($po$) to the primitive (``period 
one'') orbits; including second or higher 
repetitions does not affect the numerical results within the 
resolution of the lines presented in the figure below. As we can see 
in \fig{trace} (left panel), there exist only five ``period one'' orbits
in the system below the scaled energy $e\simeq 1.5$ corresponding to
$E\simeq 156$. 
\vspace*{-0.4cm}

\Figurebb{super}{30}{30}{770}{510}{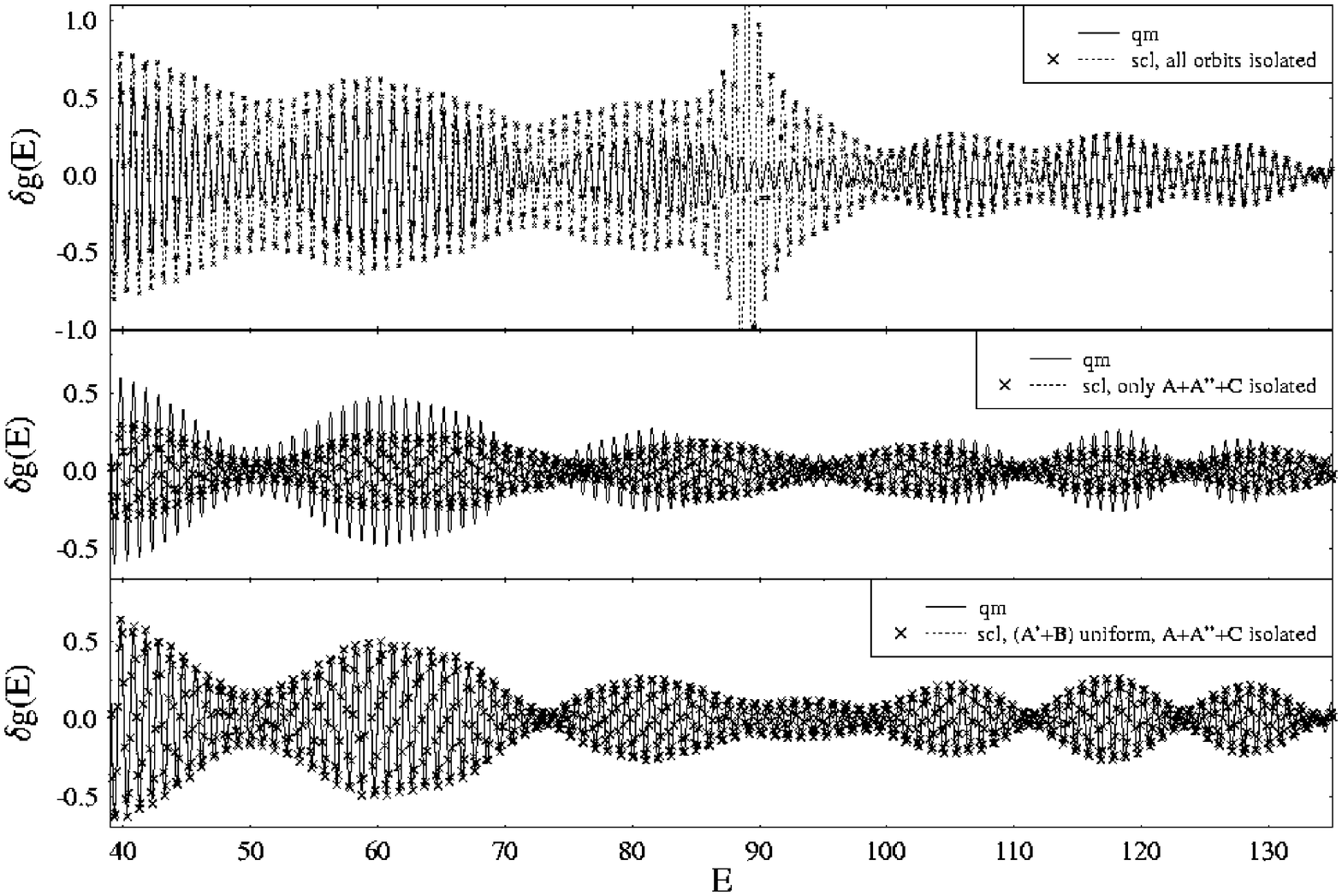}{10.8}{16.5}{
Oscillating part $\delta g(E)$ of the density of states in the GHH 
potential \eq{ghh} with $\gamma=0.6$, $\beta=0.07$ and $\alpha=0.04$, 
plotted versus the unscaled energy $E$. The quantum-mechanical (qm)
results are shown by the solid lines (identical in all three panels), 
different semiclassical (scl) approximations are shown by the crosses. 
Both qm and scl results have been coarse-grained by a Gaussian with 
width $\Delta E = 0.6$, see \eq{doscg}. 
{\it Top:} the five shortest (primitive) orbits 
A, B, C, A' and A'' are included by the standard Gutzwiller trace
formula \eq{ampiso}, \eq{tfcg} for isolated orbits. Note the divergence
near $E=89$, corresponding to $e_{tcb}=0.854447$ where the TCB of
orbits A' and B occurs (cf. \fig{cross0}). 
{\it Middle:} same as in the top panel, but the two crossing orbits A' 
and B are omitted. 
{\it Bottom:} same as in the middle panel, but now the crossing 
orbits A' and B are added in the global uniform approximation
\eq{dguni}. 
}
\vspace*{-0.4cm}

In \fig{super}, we show the oscillating part of the level density
obtained for $\alpha=0.04$ as a function of the unscaled energy $E$,
up to $E=135$ which corresponds to $e\simeq 1.3$. The solid lines 
display the coarse-grained quantum-mechanical result which is the same 
in all three panels. In order to extract the oscillating part of the 
quantum density of states \eq{doscg}, we subtracted its Strutinsky 
averaged part which here corresponds to the Thomas-Fermi approximation
\cite{book}. The crosses, connected by dashed lines, represent the 
semiclassical result in various approximations. The regular 
fast oscillations with period $\sim 1$ on the energy scale $E$ come 
from the common average action $S_{po}(E)$ of the leading periodic 
orbits, which becomes the action $S_{ho}(E)=2\pi E$ of the harmonic 
oscillator in the limit $E\to0$. The beat-like slow variation in the 
amplitude of $\delta g(E)$ is due to the interferences of the periodic 
orbits and can be captured by the semiclassical trace formula \eq{tfcg}.

In the top panel of \fig{super}, the five orbits A, B, C, A' and A'' 
are included in the trace formula \eq{tfcg} with their Gutzwiller
amplitudes \eq{ampiso}. Although they qualitatively reproduce the main 
trends of the beating density of states, they overestimate it. One can 
clearly see the divergence at $E\simeq 89$ corresponding to the scaled 
energy $e_{tcb}=0.854447$, where the TCB of the orbits A' and B occurs. 
(The divergences due to the PFB sequence of the orbit A near $e=0.993 
\leftrightarrow E=103.5$ cannot be seen with this resolution.) The center 
panel shows the same semiclassical result, but omitting the contributions 
of the bifurcating orbits A' and B. Clearly, the agreement with quantum 
mechanics is not good even far from the bifurcation, showing that these 
orbits always play a role. In the bottom panel, the orbits A, C and A'' 
are again included as isolated orbits with the amplitudes \eq{ampiso}, 
while the combined contribution of the bifurcating orbits B and A' is 
included in the global uniform approximation given in \eq{dguni}. The 
agreement between semiclassics and quantum mechanics is now excellent, 
demonstrating the adequacy of the uniform approximation. The fact that 
the isolated-orbit approximation in the top panel does not work even 
far away from the bifurcation shows that the orbits A' and B do not 
become isolated enough in the energy region shown; i.e., the asymptotic 
form \eq{dgisoab} of the uniform approximation is not reached. This 
could already be expected from the fact that the stability traces 
tr\,M$(e)$ of these orbits stay very close to +2 for all $e<1.2$, as 
can be seen in \fig{trace}.

In the energy limit $E\to 0$ (not shown in \fig{super}), the present 
semiclassical approximations are not appropriate due to the integrable 
limit of the harmonic oscillator. A corresponding uniform approximation 
for the standard HH potential has been derived in \cite{hhun}. It can 
be generalized in a straightforward manner to the GHH systems, 
following the lines of \cite{hhun}, but this would lead beyond the 
scope of the present paper.

\newpage

\section{Summary}
\label{sumc}

We have discussed transcritical bifurcations (TCBs) of periodic orbits 
in non-integrable two-dimen\-sional autonomous Hamiltonian systems. We 
have first discussed the mathematical aspects of the TCB, making use 
of recent studies by J\"anich \cite{jaen1,jaen2}. We then have, with 
the help of numerical examples in generalized H\'enon-Heiles (HH) systems,
discussed their phenomenology and their unfoldings under perturbations. 
We have shown, in particular, that a TCB may also exist in a system 
without any discrete symmetry, although it does not belong to Meyer's 
list \cite{meye} of generic bifurcations. The reason is our restriction 
to systems containing straight-line librations. In such systems, the 
TCB appears to be the generic isochronous bifurcation of the straight-line 
libration while its isochronous pitchfork bifurcation (PFB) represents 
an exception expressed by the condition $P_{qq}=0$ in \eq{extcl}. 
Using this condition and the explicit expression \eq{Pqq} for the 
calculation of $P_{qq}$ for Hamiltonians of the form \eq{ham0}, we
have exploited a special type of unfolding of the PFB to construct a 
perturbation of the standard HH system under which the TCB occurs. 

So far, we have only encountered TCBs of straight-line librational 
orbits. In most examples, the second orbit that takes part in the TCB 
with the straight-line libration is also a librational orbit, though 
not along a straight line. That this need not be so has been shown in 
the example in \fig{shapbxy}, where the second orbit is a rotation 
(except, of course, at the TCB where it coincides with the 
straight-line libration). This system, in the presence of the
momentum-dependent perturbation \eq{pertcons}, is the only example of 
a Hamiltonian which does not have the form \eq{ham0} and in 
which we found a transcritically bifurcating straight-line orbit. 
From the general 
criteria given in \Sec{seccross}, however, we see no {\it a priori} 
reason why rotational orbits should not undergo TCBs as well.
Furthermore, it is obvious that a given straight-line orbit can
always transformed into a more complicated one by suitable canonical
transformations. The inverse question -- if an arbitrary nonlinear
periodic orbit can be canonically transformed into a straight-line
libration which can bifurcate transcritically -- might also have
a positive answer, but we see no way of proving or testing this, 
nor can we give a nontrivial example, since periodic orbits (except 
straight-line librations) in non-integrable systems usually cannot 
be given analytically.

Finally, we have constructed a global uniform approximation for the 
inclusion of transcritically bifurcating periodic orbits in the 
semiclassical trace formula for the quantum density of states. A 
numerical comparison with the fully quantum-mechanical calculation 
of the coarse-grained density of states yields excellent agreement.
The normal forms of the TCB and the isochronous PFB have been derived 
in the appendix, where we also point to a ``false TCB''
which is the result of a stability exchange between different orbits
via an intermediate periodic orbit through a pair of PFBs.

\bs

\noindent
{\bf \large Acknowledgments}

\ms

\noindent
We are very grateful to K J\"anich for his vivid interest in our 
work and for critical and helpful comments to this manuscript. 
We also acknowledge encouraging discussions with J Delos, B Eckhardt, 
S Fedotkin, A Magner, J Main, M Sieber and G Tanner. 

\newpage

\section{Appendix 1: Normal forms}
\label{nofosec}


In singularity theory (see, e.g., \cite{gols}) and catastrophe theory
(see, e.g., \cite{cata}) it is standard to classi\-fy bifurcations
according to their normal forms. In \App{whynonf} we briefly discuss
normal forms for isochronous bifurcations in non-Hamiltonian
one-dimensional systems and give the explicit forms for the TCB, the
PFB and the saddle-node bifurcation (SNB).

While non-Hamiltonian fields can always be transformed to normal
forms by suitable coordinate transformations, the situation is more
complicated for Hamiltonian fields \cite{arno}. Here the normal
forms depend on pairs of canonical variables, like the Poincar\'e
variables $(q,p)$ used in \Sec{secmath}, and must be derived, for
a given Hamiltonian and a given type of bifurcating orbit, from the 
generating function ${\widehat S}(Q,p,\epsilon)$ of the Poincar\'e
map or, equivalently, from the function ${\widetilde S}(Q,p,\epsilon)$ 
given in \eq{sphase}, by suitable {\it canonical} transformations. 
The strength of the normal forms -- if they can be found -- is that 
they are unique for each generic type of bifurcation and do not depend 
on the particular form of the Hamiltonian or the bifurcating orbit. 
However, the reduction of ${\widetilde S}(Q,p,\epsilon)$ to one of 
these generic normal forms is, according to Arnold \cite{arno}, 
``generally not possible, and formal series for canonical 
transformations reducing a system to normal form generally 
{\it diverge}''. 

Nevertheless, for the generic bifurcations occurring in
two-dimensional symplectic maps, as analyzed and classified by Meyer 
\cite{meye}, normal forms suitable for semiclassical applications 
have been given in \cite{ozob,ozha}.\footnote{
We point out a misprint that occurred in both \cite{ozob} and 
\cite{ozha}: the normal form for the generic SNB was erroneously given 
analogous to that in \eq{nftcb} below, which is the normal form of the 
TCB. For the SNB the first term should correctly be $-\epsilon\,Q$ 
rather than $-\epsilon\,Q^2$.}

\subsection{Normal forms for one-dimensional non-Hamiltonian systems}
\label{whynonf}

We follow here the book of Golubitsky and Schaeffer \cite{gols} and
use their notation.
In a simple one-dimensional problem with a ``state variable'' $x$ and 
a ``bifurcation parameter'' $\lambda$, one may study the set of values 
$(x,\lambda)$ satisfying the equation
\be
g(x,\lambda) = 0\,,
\ee
where $g(x,\lambda)$ is a differentiable scalar function of both
arguments. Bifurcations of this set occur at critical points 
$(x_0,\lambda_0)$ where
\be
\left.
g_x(x_0,\lambda_0) = \papa{}{x}\,g(x,\lambda)\right|_{x_0,\lambda_0} = 0\,.
\ee
The TCB can be specified by the following criteria: at the critical
point $(x_0,\lambda_0)$, the function $g$ must fulfill 
\be
g=g_x=g_\lambda = 0\,, \qquad g_{xx}\neq 0\,, \qquad
{\rm det}\,d^2g = g_{xx}\,g_{\lambda\lambda}-g_{x\lambda}^2 < 0\,,
\label{tcbcrit}
\ee
where the subscripts denote partial derivatives with respect to the
corresponding variables. For the fixed points $x_n=x_{n+1}=x$ of the 
quadratic map \eq{quamap}, we get the function $g(x,r)=rx(1-x)-x$ 
which fulfills the criteria \eq{tcbcrit} with $\lambda=r$ at the 
critical point ($x_0=0,r_0=1$), so that a transcritical bifurcation 
must occur there. 

Normal forms are the simplest functions,
usually taken to be polynomial forms in $x$ and $\lambda$, which
obey the criteria for a given bifurcation. They can often be found
by Taylor expansion of a given equivalence class of functions
around the critical points, keeping the lowest necessary number of
terms required to fulfill the given criteria. A valid normal form for 
the TCB is \cite{guho}
\be
g_{tcb}(x,\lambda) = \pm x^2-\lambda\,x\,,
\label{htcb}
\ee
which fulfills the criteria \eq{tcbcrit} at $(x_0,\lambda_0)=(0,0)$. 
Normal forms are not unique. The following form is easily seen to be 
strongly equivalent (in the sense of \cite{gols}) to \eq{htcb}:
\be
g_{sb}(x,\lambda) = \pm x^2-\lambda^2,
\label{h1tcb}
\ee
since it also fulfills the criteria \eq{tcbcrit} at $(x_0,\lambda_0)=(0,0)$.
Although the bifurcation corresponding to \eq{h1tcb} in \cite{gols} is 
referred to as ``simple bifurcation'', it is identical to what we here
call the TCB.\footnote{Note that in \cite{gols}, the name ``transcritical'' 
is used for a whole class of bifurcations, differently from our restricted 
use of the term.}
 
The criteria for the isochronous PFB are 
\cite{gols,guho}
\be
g=g_x=g_\lambda=g_{xx}=0\,, \qquad g_{x\lambda}\neq 0\,, \qquad
g_{xxx}\neq 0\,,
\label{pfbcrit}
\ee
and its standard normal form is
\be
g_{pfb}(x,\lambda) = \pm x^3-\lambda\,x\,.
\label{hpfb}
\ee

For completeness we give here also the criteria for the SNB
(or tangent) bifurcation (in \cite{gols} called ``limit point''):
\be
g=g_x=0\,, \qquad g_\lambda\neq 0\,, \qquad g_{xx}\neq 0\,,
\label{snbcrit}
\ee
and its normal form:
\be
g_{snb}(x,\lambda)=\pm x^2-\lambda\,.
\label{hsnb}
\ee
Golubitzky and Schaeffer \cite{gols} also list the ``isola center''
bifurcation whose criteria are: 
\be
g=g_x=g_\lambda\,,\qquad g_{xx}\neq0\,,\qquad
{\rm det}\,d^2g > 0\,,
\label{iccrit}
\ee
with the normal form
\be
g_{ic}(x,\lambda) = \pm (x^2+\lambda^2)\,.
\label{hic}
\ee
In two-dimensional Hamiltonian systems, the isola center is,
according to J\"anich's classification \cite{jaen1}, a rank 1 
bifurcation for which the Hessian matrix K in \eq{Pqehess} 
is regular and definite.

\subsection{Normal forms for crossing bifurcations}
\label{nofocross}

In order to find normal forms for the two types of crossing  bifurcations 
discussed in this paper, which we need in the semiclassical trace formula 
\eq{phsptf}, we follow the heuristic approach of determining a truncated 
Taylor expansion of ${\widetilde S}(Q,p,\epsilon)$ with the minimum 
number of terms necessary to describe the required properties of these 
bifurcations. To this purpose, we first establish relations between
the partial derivatives of the functions $Q(q,p,\epsilon)$ and 
$P(q,p,\epsilon)$ in \eq{QPofqp}, and the partial derivatives of the 
function ${\widetilde S}(Q,p,\epsilon)$ for which we use the same notation 
as in \ref{secappsm}. Translating the criteria given in \Sec{seccross} 
for the crossing bifurcations in terms of the partial derivatives of 
${\widetilde S}$, we can determine the normal forms of the TCB and 
the FLB. Although the formal transformations needed to arrive at
these normal forms are not necessarily canonical, their use in the
semiclassical trace formula can be justified by the fact that
possible missing terms of higher order do not affect the results to 
leading order in $\hbar$ (cf.\ \cite{ozob,ssu1,sun,ssu2}).

\newpage

\subsubsection{Relations between $Q,P$ and $\widetilde S$ and their
partial derivatives}

From \eq{sphase} and \eq{canonrel} we obtain the following basic
relations
\bea
Q(q,p,\epsilon) & = & q - {\widetilde S}_p(Q,p,\epsilon)\,,\nonumber\\
P(q,p,\epsilon) & = & p + {\widetilde S}_Q(Q,p,\epsilon)\,.
\label{QPtoStil}
\eea
We now take partial derivatives of these relations with respect to the 
variables $q$, $p$ and $\epsilon$, in order to formulate the conditions 
for the crossing bifurcations discussed in \Sec{seccross} in terms of
partial derivatives of the function ${\widetilde S}(Q,p,\epsilon)$.
This procedure is simplified by the following step. The {\bf splitting 
lemma} of catastrophe theory (see \cite{cata}, pp 95 and 103) states 
that after a suitable (but perhaps not canonical) coordinate 
transformation, the function ${\widetilde S}(Q,p,\epsilon)$ can be 
split up in the following way 
\be
{\widetilde S}(Q,p,\epsilon) = S(Q,\epsilon) - \frac{\sigma}{2}\,p^2,
\qquad \sigma\neq 0\,, 
\label{split}
\ee
where $S(Q,\epsilon)$ does not depend on $p$ any more. In the suitably 
adapted coordinates $q,p$ and $Q,P$ for which \eq{split} is
true,\footnote{and for which $\MA(0)$ has the form \eq{trma0l} [see 
also the footnote after \eq{trma0l}]} we obtain the following relations
\bea
& \!\!\!\!\!\!\!\!\!\!\!\!\!\!\!Q_q(\epsilon) = 1\,, 
& \quad Q_p(\epsilon) = -{\widetilde S}_{pp}(\epsilon)= \sigma\,,\nonumber\\ 
& P_q(\epsilon) \,= S_{QQ}(\epsilon)\,,
& \quad P_p(\epsilon) \,= 1+\sigma\,S_{QQ}(\epsilon)\,, 
\label{partder}
\eea
and $\trM(\epsilon)$ becomes
\be
\trM(\epsilon) = 2 + \sigma\,S_{QQ}(\epsilon)\,,
\label{trmagen}
\ee
which is valid along the fixed-point branches of both orbits A and B.
We also give some of the higher partial derivatives of $Q$ and $P$
at $\epsilon=0$ (valid in the adapted coordinates):
\be
Q_{qq} = 0\,,       \quad  Q_{qp} = 0\,, \quad
P_{qq} = S_{QQQ}\,, \quad  P_{qp} = \sigma\,S_{QQQ}\,, 
\ee
and
\be
P_{q\epsilon} = S_{QQ\epsilon}\,, \quad P_{qqq} = S_{QQQQ}\,.
\ee

\subsubsection{Criteria for the two crossing bifurcations}

We can now express the criteria for the two types of crossing 
bifurcations introduced in \Sec{seccross} directly in terms 
of the parameter $\sigma$ and the partial derivatives of the 
function $S(Q,\epsilon)$ defined in \eq{split}.
To have a bifurcation of the A orbit at $\epsilon=0$, we must have
\be
S_{QQ}(Q=0,\epsilon=0) = 0\,. \qquad (\hbox{\it bifurcation~of~A~orbit})
\label{SQQ0}
\ee
For the occurrence of a rank 1 bifurcation, we have the criterion
(see the end of \Sec{secappsm})
\be
P_\epsilon = S_{Q\epsilon} = 0\,. \qquad (\hbox{\it rank~1~bifurcation})
\label{SQeps}
\ee
Since $(P,Q)=(p,q)=(0,0)$ then is the fixed-point branch of the A orbit 
for all $\epsilon$, the function $S(Q,\epsilon)$ must fulfill, due to 
\eq{QPtoStil}, the condition
\be
S_Q(0,\epsilon) = 0 \qquad \forall \,\epsilon\,. 
                  \qquad (\hbox{\it fixed-point~branch~of~A~orbit})
\label{SQ}
\ee

The criterion for the occurrence of a crossing bifurcation 
is that the slope $\trMA'(0)$ given by \eq{slopea} be nonzero,
see also \eq{Pqenot0}. We therefore need
\be
S_{pp}= -\sigma \neq 0\,, \quad S_{QQ\epsilon} \neq 0\,. 
                          \qquad (\hbox{\it  crossing~bifurcation})
\label{crossb}
\ee
The criterion for this bifurcation to be transcritical is
\be
S_{QQQ} \neq 0\,. \qquad (\hbox{\it transcritical~bifurcation})
\label{transb}
\ee
For the occurrence of a fork-like bifurcation, we must have
\be
S_{QQQ} = 0\,, \qquad S_{QQQQ} \neq 0\,.  
                          \qquad (\hbox{\it fork-like~bifurcation})
\label{forkb}
\ee

We are now ready to construct the simplest normal forms for the
function ${\widetilde S}(Q,p,\epsilon)$, split like in \eq{split}, 
that fulfill all the above criteria \eq{SQQ0} - \eq{crossb} and 
either \eq{transb} or \eq{forkb}.

\subsubsection{Normal form of the TCB}
\label{secnormtcb}

For the {\bf transcritical bifurcation}, the normal form obtained in
this way is
\be
{\widetilde S}(Q,p,\epsilon)
                 = -\,\epsilon\,Q^2 - b\,Q^3 - \frac{\sigma}{2}\,p^2\,, 
                      \qquad (b\neq 0)
\label{nftcb}
\ee
with $b=-\frac16\,P_{qq}$. An explicit formula for calculating 
$P_{qq}$ and hence the parameter $b$ is given in \eq{Pqq}. The normal 
form \eq{nftcb} corresponds to that of the TCB in non-Hamiltonian 
systems given in \eq{htcb} in \App{whynonf}, if we choose 
${\widetilde S}_Q(Q,p$=$0,\epsilon)=g_{tcb}(q,\epsilon)$, but 
it has, to our knowledge, not been discussed in connection with 
bifurcations of periodic orbits in Hamiltonian systems.

The fixed-point branch of the B orbit is easily found to be
\be
p_B(\epsilon) = 0\,; \qquad
Q_B(\epsilon) = -\frac{2}{3b}\,\epsilon \quad \Leftrightarrow  \quad
\epsilon_B(Q) = -\frac{3b}{2}\,Q\,.
\ee
The stability traces of the two orbits are then found from
\eq{trmagen} to be
\bea
\trMA(\epsilon)  & = & 2-2\sigma\epsilon\,,\nonumber\\
\trM_{\rm B}(\epsilon) & = & 2+2\sigma\epsilon\,,
\label{trMn}
\eea
fulfilling the ``TCB slope theorem'' \eq{slopestc}. Along the branch $B$, 
the function ${\widetilde S}(Q_B,p_B,\epsilon)$ yields a contribution 
to the action of the B orbit. Noting that the contribution to the A 
orbit is zero, this yields the {\bf action difference} of the two orbits:
\be
{\widetilde S}(Q_B,p_B,\epsilon) = \Delta S 
           = S_{\rm B}-S_{\rm A} = - \frac{\epsilon^3}{6b^2}\,.
\label{delSntcb}
\ee

Note that a sign change of either $\sigma$ or $\epsilon$ in \eq{nftcb}
simply corresponds to exchanging the orbits A and B, whereas a sign 
change of $b$ does not affect the local predictions \eq{trMn} and
\eq{delSntcb}.

In the applications of the normal forms for semiclassical uniform
approximations, one usually assumes $\sigma=\pm1$ (see, e.g., 
\cite{sun,ssu1}). However, when starting from an arbitrary 
Hamiltonian, this is not automatically fulfilled. In fact, one 
sees directly from \eq{partder} that $\sigma=Q_p$ which {\it a
priori} is not of modulus unity. But we can easily absorb the 
absolute value of $\sigma$ by a canonical stretching (shear) 
transformation: $(Q,p)\rightarrow ({\widetilde Q}\,,\,{\tilde p})$ 
specified by
\be
\widetilde{Q} = Q/\!\sqrt{|\sigma|}\,, \qquad
\widetilde{p} = p\,\sqrt{|\sigma|}\,.
\ee
The normal form \eq{nftcb} then becomes
\be
{\widetilde S}({\widetilde Q},{\tilde p},{\tilde\epsilon})
                      = -\,{\tilde \epsilon}\,{\widetilde Q}^2 
                        - {\widetilde b}\,{\widetilde Q}^3 
                        - \frac{{\tilde\sigma}}{2}\,{\tilde p}^2,\qquad 
                          {\tilde\sigma} = \pm 1
\label{nftcbtil}
\ee
with
\be
{\tilde\epsilon} = |\sigma|\epsilon\,,\qquad
{\widetilde b} = |\sigma|^{3/2}\,b\,.
\ee
In \Sec{secunifo} we shall use the form \eq{nftcbtil} but omit the tilde 
on all variables and constants.

\subsubsection{Normal forms for two unfoldings of the TCB}
\label{secnormunftcb}

We have found two scenarios for the destruction of a TCB by a
perturbation $\kappa\,\delta H(x,y,p_x,p_y)$ of the Hamiltonian, where 
$\kappa$ is a real parameter. In the first scenario, the bifurcation 
unfolds into a pair of SNBs lying opposite to each 
other on either side of the unperturbed bifurcation point $\epsilon$. 
This scenario can be described by adding to the normal form \eq{nftcb} 
a term linear in $Q$ with a negative sign:
\be
{\widetilde S}(Q,p,\epsilon) =
    -\,\kappa^2Q -\,\epsilon\,Q^2 - b\,Q^3 - \frac{\sigma}{2}\,p^2. 
     \qquad (b>0)
\label{nf2tb}
\ee
It predicts the following local behavior of the stability traces: 
\be
\hbox{tr}\,{\rm M}_{\rm A,B} = 2 \pm 2\sigma\sqrt{\epsilon^2-3b\,\kappa^2}\,.
\label{trm2tb}
\ee
Between the two SNBs, which occur at 
$\epsilon=\pm\sqrt{3b}\,\kappa$, there are no real periodic orbits. 
For $\epsilon<-\sqrt{3b}\,\kappa$ and for $\epsilon>
\sqrt{3b}\,\kappa$, the pairs of original orbits A and B join and
``destroy'' each other in the SNBs. Examples for this
scenario are given in \Sec{secbfield}, where $\kappa$ is proportional 
to the strength of a homogeneous external magnetic field, and in 
\Sec{secdespot}.

In the second scenario, no bifurcation is left in the presence of the
perturbation. The two pairs of orbits approach the critical line
tr\,M = 2 from both sides, come closest to it at the original 
bifurcation point $\epsilon=0$, and then diverge from it again. We
will call this an ``avoided bifurcation''. It can be described by
a normal form identical to \eq{nf2tb}, except for an opposite sign of
the linear term:
\be
{\widetilde S}(Q,p,\epsilon) =
    +\,\kappa^2Q -\,\epsilon\,Q^2 - b\,Q^3 - \frac{\sigma}{2}\,p^2. 
    \qquad (b>0)
\label{nfavoid}
\ee
This form predicts for the local behavior of the stability traces 
\be
\hbox{tr}\,{\rm M}_{\rm A,B} = 2 \pm 2\sigma\sqrt{\epsilon^2+3b\,\kappa^2}\,,
\label{trmavoid}
\ee
corresponding to an avoided bifurcation. An example of this unfolding of
the TCB is given in \Sec{secavoid}.

Both types of unfoldings have been found also in non-Hamiltonian
one-dimensional systems (see, e.g., \cite{bad1}). According to \cite{gols}, 
one may describe a ``universal unfolding'' of the TCB by the normal form
\be
{\widetilde S}(Q,p,\epsilon) = \delta\,Q - \epsilon\,Q^2 - b\,Q^3 
                             - \frac{\sigma}{2}\,p^2. \qquad (b>0)
\label{unvunftcb}
\ee
This corresponds to \eq{nf2tb} for $\delta<0$ and to \eq{nfavoid} for
$\delta>0$. The above two normal forms are, however, mathematically
different in that they predict only one type of unfolding for both
signs of the parameter $\kappa$ in each case, whereas \eq{unvunftcb}
would predict different unfoldings for the two signs of one and the
same parameter $\delta$. We emphasize that all three cases describe 
different (physical) phenomena; we have not found any realization of 
the type \eq{unvunftcb} in our numerical studies of TCBs.

\subsubsection{Normal form of the FLB}
\label{secnormpfb}

For the {\bf fork-like bifurcation} (FLB), we arrive at the normal form
\be
{\widetilde S}(Q,p,\epsilon)
                 =  -\,\epsilon\,Q^2 - a\,Q^4  - \frac{\sigma}{2}\,p^2, 
                      \qquad (a\neq 0)
\label{nfpfb}
\ee
with $a=-\frac{1}{24}\,P_{qqq}$. This form is 
identical to that of the generic period-doubling PFB \cite{ozob} 
and corresponds, with ${\widetilde S}_Q(Q,p$=$0,\epsilon)=
g_{pfb}(q,\epsilon)$, to that of the isochronous PFB in
non-Hamiltonian systems given in \eq{hpfb}.

The two fixed-point branches of the B orbits here are given locally by
\be
p_B(\epsilon) = 0\,; \qquad
Q_B(\epsilon) = \pm\sqrt{-\epsilon/2a} \quad \Leftrightarrow  \quad
\epsilon_B(Q) = -2a\,Q^2\,,
\ee
where the rightmost relation fulfills the conditions $\epsilon_B'(0)=0$, 
$\epsilon_B''(0)\neq 0$ given in \eq{Bbranch}. The stability traces of 
the A and B orbits (on that side of $\epsilon=0$ where the latter
exist) are found locally to be
\bea
\trMA(\epsilon)  & = & 2-2\sigma\epsilon\,,\nonumber\\
\trM_B(\epsilon) & = & 2+4\sigma\epsilon\,,
\label{trMnpf}
\eea
fulfilling the ``FLB slope theorem'' \eq{slopesfb}, and their action 
difference becomes
\be
{\widetilde S}(Q_B,p_B,\epsilon) = \Delta S 
           = S_{\rm B}-S_{\rm A} = \frac{\epsilon^2}{4a}\,.
\label{delSnpfb}
\ee
The same local behaviors \eq{trMnpf} and \eq{delSnpfb} hold also
for the generic PFB \cite{ssu1}.

Note that the B branches describing the bifurcated new orbits B only 
exist on that side of the bifurcation where $\epsilon/a<0$. Changing the
sign of $\epsilon$ has the same effect as changing the sign of $a$. 
For the GHH potentials, all isochronous PFBs of the A type orbits have 
negative values of $a$.

Changing the sign of $\sigma$ ``mirrors'' the bifurcation scenario 
at the line $\trM=+2$, i.e., the stabilities of all orbits are exchanged 
from stable to unstable and vice versa.

\subsubsection{Normal form for unfoldings of the FLB}
\label{secnormunfpfb}

We have found two scenarios for unfoldings of the FLB, which 
are known also to occur in non-Hamiltonian systems 
\cite{bad1,gols}. Since the FLB locally is identical to the
isochronous PFB, one may use its universal unfolding in
one-dimensional non-Hamiltonian systems \cite{gols} as a 
guide and choose the normal form
\be
{\widetilde S}(Q,p,\epsilon)
       =  -\,\epsilon\,Q^2 - a\,Q^4 + \delta\,Q + \kappa\,Q^3 
          - \frac{\sigma}{2}\,p^2,  \qquad (a\neq 0)
\label{unfpfb}
\ee
thus adding a linear and a cubic term in $Q$ to \eq{nfpfb}. One 
type of unfolding is obtained for $\delta\neq0$. Hereby a pair of 
orbits is created in a SNB on one side of a critical value 
$\epsilon_1$ of $\epsilon$ (depending on $\delta$ and $\kappa$), 
while a third orbit is present on both sides of $\epsilon$ and does 
not undergo any bifurcation. This is the most usual unfolding of 
the FLB (and persistent in the sense of \cite{gols}). Examples are 
given in \fig{etafig} (inserted close-up) and in \fig{gamfig}.

The second scenario, more interesting for our present investigations,
is obtained from \eq{unfpfb} as the special case $\delta=0$. Hereby 
the FLB (or isochronous PFB) is broken up into a TCB, occurring at 
the original bifurcation point $\epsilon=0$, associated with a 
near-lying SNB. Examples of this unfolding are shown in \fig{trace} 
(left panel) and in \fig{bifl6}. The fixed-point set in this 
unfolding is found from ${\widetilde S}_p=p=0$ and 
\be
{\widetilde S}_Q(Q,\epsilon)=-2\epsilon\,Q-4a\,Q^3-3\kappa\,Q^2=0\,,
\label{unfpfb0}
\ee
yielding the fixed-point branch $(Q_A,Q_B)=(0,0)$ of the original A 
orbit, which is not affected by the perturbation with $\kappa\neq0$, 
and a set of new points $(Q_B,p_B=0)$, where $Q_B$ are solutions of
the equation
\be
4a\,Q^2+3\kappa\,Q+2\epsilon=0\,.
\label{tiltparab}
\ee
This corresponds to a parabola in the ($Q,\epsilon$) plane.
Calculating the stability traces tr\,M$_A(\epsilon)$ and 
tr\,M$_B(\epsilon)$, one obtains the graphs seen in \fig{bifl6}.
The A branch is locally linear around $\epsilon=0$. The B branches,
realized here by the L$'_6$, L$_7$ and L$_6$ orbits, form a tilted 
parabola with the following properties.
($i$) The symmetry line of the parabola has the slope tr\,M$'_B(0)$ 
of the original B orbits created at the unperturbed FLB, obeying 
the ``FLB slope theorem'' tr\,M$'_B(0)$ = -2\,tr\,M$'_A(0)$ given in
\eq{slopesfb}. 
($ii$) The parabola intersects the A orbit in the new TCB with
the slope $-$tr\,M$'_A(0)$ according to the ``TCB slope theorem'' 
\eq{slopestc}.

\newpage

\section{Appendix 2: False transcritical bifurcations}
\label{falsetcb}

We briefly address here a mechanism of stability exchange which 
has been described in \cite{erda}. Hereby a pair of periodic orbits 
exchange their stabilities via an intermediate periodic orbit which 
is exchanged between the two other orbits through two oppositely
oriented isochronous PFBs. On a moderate scale of the control 
parameter, the intermediate orbit may not be observed numerically, 
and the stability traces of the two main orbits appear to cross the 
critical line tr\,M = 2 exactly like in a TCB. These are, however, 
``false transcritical bifurcations''.

We illustrate this in \fig{stabex}, calculated for the
coupled quartic oscillator potential $V(x,y)=(x^4+y^4)/4
+\alpha\,x^2y^2\!/2$ which is non-integrable for all values of 
$\alpha$ except $\alpha=0$, 1, and 3 (see \cite{lame,erda,bfmm}). 
Shown are the stability traces tr\,M$(\alpha)$ of two orbits P 
and F which are created from a period-tripling bifurcation of 
a straight-line orbit at $\alpha=0.6315$ and exist only for 
$\alpha\leq 0.6315$. They are isolated at all values of $\alpha$
except for $\alpha=0$. At $\alpha=0$ and near $\alpha\sim 0.0055$ 
they appear to cross like in a TCB; note that also the same 
exchange of Maslov indices by one unit takes place for each 
orbit. 

\Figurebb{stabex}{10}{10}{767}{260}{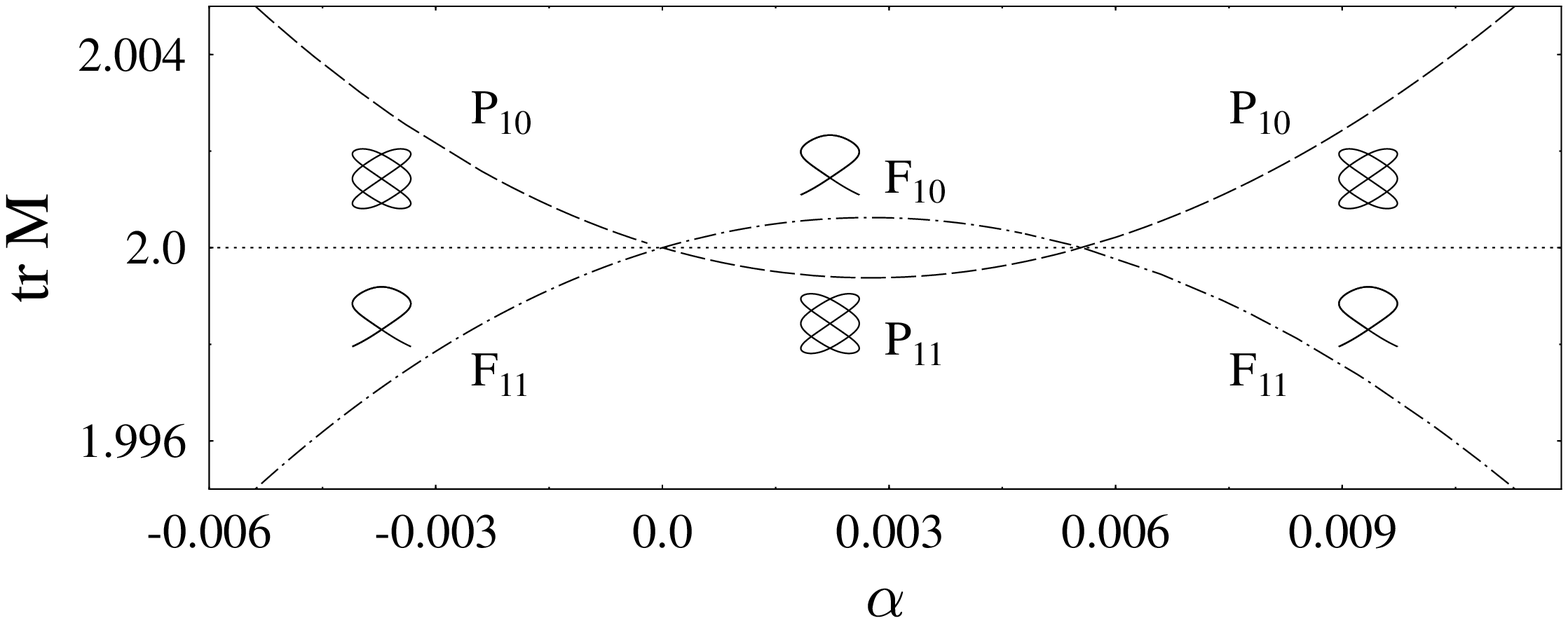}{5}{16.5}{
Stability exchanges of orbits F and P in the coupled
quartic oscillator. Both crossings are {\bf not} 
transcritical bifurcations. (See text for details.)}
\vspace*{-0.5cm}

However, the two orbits have different shapes at all values of 
$\alpha$, as shown by the inserts which display their shapes 
in the $(x,y)$ plane. Therefore, their fixed points cannot coincide 
at either of the crossings, and hence the crossing near $\alpha\sim 
0.0055$ {\bf cannot} be a TCB. The situation around $\alpha=0$, where 
the system is integrable, is not a bifurcation at all, but the 
generic Poincar\'e-Birkhoff breaking of a rational (3:2) torus into 
a pair of stable and unstable isolated orbits.

What actually happens near $\alpha\sim 0.0055$, as described
in \cite{erda}, is shown in \fig{stabexz}, where the scale
of the graphs tr\,M($\alpha$) has been enlarged in both 
directions by a factor $\sim 10^{5}$. 
The graphs tr\,M($\alpha$) really cross slightly above the 
critical line tr\,M = 2 (note the shift along the vertical axis
by two units), here at a distance of only $\sim 10^{-7}$
which requires a high precision of the numerical calculations.
Their crossings of the critical line, where
bifurcations {\bf must} take place, occurs at two different 
points in a tiny distance $\Delta\alpha=3.56\times10^{-7}$.
These bifurcations are isochronous PFBs, and the orbit
Q emitted and reabsorbed by them intermediates between the
shapes of the two crossing orbits. Note that the orbit Q,
which transforms the shape of the F orbit into that of the
P orbit (or {\it vice versa}), has twice the discrete
degeneracy of the orbits F and P. Compared to F this is
because Q is a rotation and has two time orientations (while
F transforms into itself under time reversal); compared to
P it has a lower discrete symmetry (it is not symmetric under
reflection at the $x$ axis, but P is). This double degeneracy
with respect to the parent orbits is characteristic of the
isochronous (non-generic) PFB (see \Sec{secpfb}). 

The change of the shape of the intermediate
orbit Q is very similar to that of an orbit on an integrable
torus -- and can hardly be distinguished numerically from it.
(It is exactly the case in the nearby integrable point 
$\alpha=0$, where P and F cross again and correspond to two
realizations of one and the same 3:2 torus.) Under poor 
numerical resolution -- as in \fig{stabex} above -- therefore,
one might also misinterpret the situation at $\alpha\sim0.0055$
as existence of a locally integrable torus.

\Figurebb{stabexz}{10}{10}{767}{470}{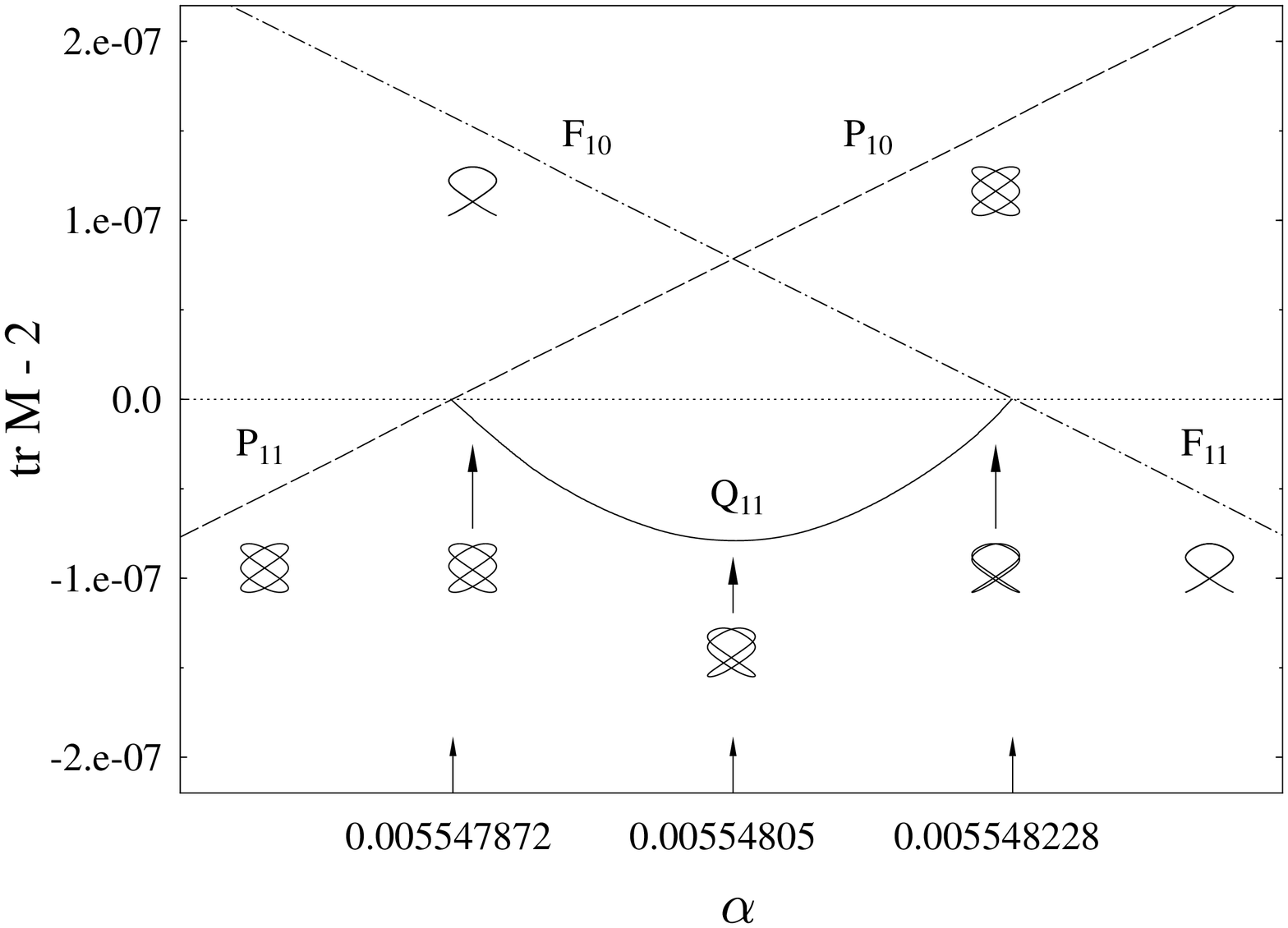}{8.5}{16.5}{
Same as \fig{stabex}, but the scales in both directions
are enlarged by a factor $\sim 10^{5}$. Note how the
orbit Q intermediates between the orbits F and P
through two pitchfork bifurcations.}

Although this mechanism has been observed and published quite
a long ago \cite{erda}, it does not appear to be widely known.
We deem it worth mentioning and illustrating here, in order 
to prevent misinterpretations of such ``false'' TCBs that can 
appear under poor numerical circumstances. (Although misunderstanding 
should not happen when the shapes of the orbits are 
known and/or when it is realized that their fixed points on the 
Poincar\'e surface of section do not coincide at the crossing.)

\end{document}